\documentclass[journal]{IEEEtran}
\usepackage[T1]{fontenc}
\usepackage{times}
\usepackage{dcolumn}
\usepackage{endnotes}
\usepackage{graphics}
\usepackage{times}
\usepackage{epsfig}
\usepackage{graphicx}
\usepackage{textcomp}
\usepackage{amssymb,amsmath}
\usepackage{balance}
\usepackage{float}
\usepackage[noadjust]{cite}

\usepackage{listings}
\usepackage[]{algorithm2e}
\usepackage{flushend}
\usepackage{soul,color}
\usepackage[colorlinks=true,linkcolor=black,anchorcolor=black,citecolor=black,filecolor=black,menucolor=black,runcolor=black,urlcolor=black]{hyperref}
\usepackage{verbatim} 
\usepackage{wrapfig}
\usepackage{siunitx}
\usepackage{todonotes}
\usepackage{eucal}
\usepackage[cal=dutchcal]{mathalfa}
\usepackage{bm}
\usepackage{makecell} 
\usepackage{pifont}
\usepackage{blindtext}%
\definecolor{cadmiumgreen}{rgb}{0.0, 0.42, 0.24}

\soulregister\cite7
\soulregister\ref7
\makeatletter
\newcount\SOUL@minus 
\makeatother
\newcommand{\xmark}{\ding{55}}%
\newcommand{\cmark}{\ding{51}}%

\begin{document}

\title{
Vector Symbolic Architectures as a Computing Framework for Emerging Hardware
}

\author{
Denis~Kleyko,
Mike~Davies,
E.~Paxon~Frady, 
Pentti~Kanerva,
Spencer~J.~Kent, 
Bruno~A.~Olshausen, 
Evgeny~Osipov, 
Jan~M.~Rabaey, 
Dmitri~A.~Rachkovskij, 
Abbas~Rahimi, and 
Friedrich~T.~Sommer 
\thanks{Manuscript received on 4 June 2021; revised on 11 April 2022 and 30 June 2022; accepted 2 September 2022.
DK has received funding from the European Union's Horizon 2020 research and innovation programme under the Marie Sk\l{}odowska-Curie grant agreement No 839179.
The work of BAO, JMR, PK, and DK  was supported in part by the DARPA's VIP (Super-HD Project) and AIE (HyDDENN Project) programs.
The work of DK, PK, and BAO  was supported in part by AFOSR FA9550-19-1-0241.
The work of FTS, BAO, and DK  was supported in part by Intel's THWAI program.
The work of DAR was supported in part by the National Academy of Sciences of Ukraine (grant no. 0120U000122, 0121U000016, 0122U002151, and 0117U002286), the Ministry of Education and Science of Ukraine (grant no. 0121U000228 and 0122U000818), and the Swedish Foundation for Strategic Research (SSF, grant no. UKR22-0024).
FTS was supported by NIH R01-EB026955. 
} 
\thanks{D. Kleyko is with the Redwood Center for Theoretical Neuroscience at the University of California at Berkeley, CA 94720, USA and also with the Intelligent Systems Lab at Research Institutes of Sweden, 16440 Kista, Sweden. \mbox{E-mail}: \mbox{denis.kleyko@ri.se}
}
\thanks{M. Davies and E. P. Frady are with the Neuromorphic Computing Laboratory, Intel Labs, Santa Clara, CA 95054, USA.
\mbox{E-mail}: \mbox{mike.davies@intel.com, e.paxon.frady@intel.com}
}
\thanks{P. Kanerva, S. J. Kent, and B. A. Olshausen are with the Redwood Center for Theoretical Neuroscience at the University of California at Berkeley, CA 94720, USA. \mbox{E-mail}: \mbox{pkanerva@berkeley.edu, spencer.kent@berkeley.edu,} \mbox{baolshausen@berkeley.edu}%
}
\thanks{E. Osipov is with the Department of Computer  Science Electrical and Space Engineering, Lule\aa{} University of Technology, 97187 Lule\aa{}, Sweden. \mbox{E-mail}: \mbox{evgeny.osipov@ltu.se}
}
\thanks{J. M. Rabaey is with the Department of Electrical Engineering and Computer Sciences at the University of California at Berkeley, CA 94720, USA. \mbox{E-mail}: \mbox{jan\_rabaey@berkeley.edu}
}
\thanks{D. A. Rachkovskij is with International Research and Training Center for Information Technologies and Systems, 03680 Kyiv, Ukraine, and with the Department of Computer  Science Electrical and Space Engineering, Lule\aa{} University of Technology, 97187 Lule\aa{}, Sweden.
\mbox{E-mail}: \mbox{dar@infrm.kiev.ua}
 }
\thanks{A. Rahimi is with IBM Research -- Zurich, 8803 R\"{u}schlikon,  Switzerland. \mbox{E-mail}: \mbox{abr@zurich.ibm.com}
}
\thanks{F. T. Sommer is with the Redwood Center for Theoretical Neuroscience at the University of California at Berkeley, CA 94720, USA and also with the Neuromorphic Computing Laboratory, Intel Labs, Santa Clara, CA 95054, USA. \mbox{E-mail}: \mbox{fsommer@berkeley.edu}%
}
 }

\markboth{PROCEEDINGS OF THE IEEE}%
{Kleyko \MakeLowercase{\textit{et al.}}: Vector Symbolic Architectures}

\maketitle


\begin{abstract}
This article reviews recent progress in the development of the computing framework {\em Vector Symbolic Architectures} (also known as Hyperdimensional  Computing). This framework is well suited for implementation in stochastic, emerging hardware and it naturally expresses the types of cognitive operations required for Artificial Intelligence (AI).
We demonstrate in this article that the field-like algebraic structure of Vector Symbolic Architectures offers simple but powerful operations on high-dimensional vectors that can support all data structures and manipulations relevant to modern computing. 
In addition, we illustrate the distinguishing feature of Vector Symbolic Architectures,  ``computing in superposition,'' which sets it apart from conventional computing. 
It also opens the door to efficient solutions to the difficult combinatorial search problems inherent in AI applications.
We sketch ways of demonstrating that Vector Symbolic Architectures are computationally universal. 
We see them acting as a framework for computing with distributed representations that can play a role of an abstraction layer for emerging computing hardware.
This article serves as a reference for computer architects by illustrating the philosophy behind Vector Symbolic Architectures, techniques of distributed computing with them, and their relevance to emerging computing hardware, such as neuromorphic computing.

\end{abstract}


\section{Introduction}
\label{sect:intro}

The demands of computation are changing. 
First, Artificial Intelligence (AI) and other novel applications pose a host of computing problems that require a search over an immense space of possible solutions, with many approximately correct answers, but rarely a single correct one. 
Second, future emerging hardware platforms, operating at ultra-low voltages to reduce energy consumption and to support continued process scaling, are  destined to be noisy and, hence, operate stochastically~\cite{JaegerComputing2020}.
These observations expose the need for a computing framework that supports both deterministic computation in the presence of noise as well as the approximate and parallel nature of algorithms used in AI.

By emerging hardware, we refer to the broad class of new hardware designs that are highly parallel, fabricated at ultra-small scales, utilize novel components, and/or operate at ultra-low voltages, thus consisting of unreliable, stochastic computational elements.

The conventional (\`{a} la von Neumann) computing architecture is not well adapted to these demands, as it was designed assuming precise computational elements for tasks that require exact answers.  
Conventional computing architectures will continue to play an important role in technology, but there is a growing amount of computational demands that are better served by new computing designs.
Thus, hardware engineers have been looking at distributed and neuromorphic computing as a way of meeting these demands.

Many of the emerging computational demands come from cognitive or perceptual applications found within the realm of AI.
Examples include image recognition, computer vision, and text analysis.
Indeed, large-scale deep learning neural network modeling dominates discussions about modern computing technology, pushing innovations in hardware design towards parallel, distributed processing~\cite{ben2019demystifying}.
While widely used, deep learning neural networks still have limitations, such as lacking the transparency of learned representations and the difficulties in performing symbolic computations. 
In order to support more sophisticated symbolic computations, researchers have been embedding conventional data structures, such as graphs and key-value pairs, into neural network models~\cite{KipfS-SGCNN2016, ScarselliGNN2008, VaswaniAttention2017}. 
However, it is not yet clear whether the sub-symbolic pattern recognition and learning capabilities of deep neural networks can be augmented to handle the rich control flow, abstraction, symbol manipulation, and recursion of existing computing frameworks.

Work on developing emerging computing hardware is accelerating.
There are many showcase demonstrations~\cite{LinOpticalDNN2018, PeiTianjic2019,ImamOlfactory2020, DaviesAdvancingLoihi2021} but so far:
\begin{itemize}
    \item these demonstrations have mostly lacked a unifying theoretical framework that can bring sufficient composability, explainability, and versatility;
    \item many demonstrations still depend on hand-crafted elements that would be prone to errors;
    \item most of the demonstrations have been sub-symbolic in nature and resort to support from the conventional computing architecture to implement the symbolic and flow control elements.
\end{itemize}
While these points are valid in general, there are some exceptions which we discuss in Section~\ref{sec:related:framework}. 
Nevertheless, all of these demonstrate the need for a unifying computing framework that can serve as 
an abstraction layer between hardware and desired functionality.
Ideally, such a framework should be flexible enough to provide interfaces to emerging hardware with various  features, such as stochastic components, asynchronous spiking communication, or devices with analog elements.

\begin{figure}[t]
\centering
\includegraphics[width=0.75\columnwidth]{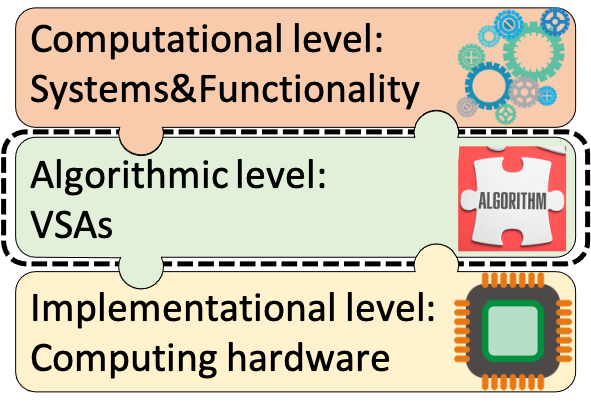}
\caption{
The place of HDC/VSA within Marr's levels of analysis \cite{MarrVision1982}. The focus of this article is marked by the dashed rectangle. We explain how HDC/VSA provides primitives to formalize algorithms in ways that seamlessly connect to the computational and implementational levels in the computing hierarchy.
}
\label{fig:stack}
\end{figure}

For the following reasons, we propose Vector Symbolic Architectures (VSA)~\cite{Gayler2003} or, synonymously, Hyperdimensional Computing (HDC)~\cite{Kanerva09} as such a computing framework. 
First, HDC/VSA can represent and manipulate both symbolic and numerical data structures with distributed vector representations 
to solve, e.g., 
cognitive~\cite{EliasmithSPAUN2012, RachkovskijSimilarity2012, EmruliAnalogical2013} or machine learning~\cite{HDClassGe} tasks.
HDC/VSA is a suitable framework for integration with neural network computations for solving problems in AI. 
It extends beyond typical AI tasks 
as an approach capable of performing symbolic manipulations with distributed representations. 
Second, the design of HDC/VSA, which was inspired by the brain, lends itself to implementation in emerging computing technologies~\cite{MemristorHD19} because it is highly robust to individual device variations.
Third, HDC/VSA is a framework with two interfaces, one towards computations and algorithms and one towards implementation and representations (cf. Fig.~\ref{fig:stack}). There are different HDC/VSA models that all offer the same operation primitives but differ slightly in terms of their implementation of these primitives. For example, there are HDC/VSA models that compute with binary, bipolar, continuous real, and continuous complex vectors. Thus, the HDC/VSA concept has the flexibility to connect to a multitude of different hardware types, such as analog in-memory computing architectures~\cite{MemristorHD19} for binary-valued HDC/VSA models  or spiking neuron architectures~\cite{RennerBinding2022,BentSpike2022} for complex-valued ones.

HDC/VSA is a relatively new concept. 
The key idea goes back to the 1990s, but computers of the day were not ready to handle large numbers of high-dimensional vectors. 
Now they are, and so the framework deserves to be looked into anew. 
Not as a complete substitute for conventional computing, but as a concept complementing it in a specific niche. 
For example, human and animal-like perception and learning have eluded our attempts to be programmed into computers. 
HDC/VSA is a strong candidate for such tasks because of their suitability for both statistical learning and symbolic reasoning.

This article provides three main contributions. 
First, we review the principles of HDC/VSA and how they provide a generic computing framework for implementing the primitives of conventional data structures and deterministic algorithms. 
Second, we highlight pros and cons of a non-traditional mode of computing in HDC/VSA, ``computing in superposition,'' which can leverage distributed representations and parallelism for efficiently solving computationally hard problems. 
Finally, we present two proposals (see Appendix~\ref{sect:turing}) that show the universality of HDC/VSA by using them to represent systems known to be Turing complete.

\subsection*{Guide to the article}

The article is written with both newcomers to HDC/VSA and seasoned readers in mind. 
Section~\ref{sect:motiv} provides some motivation for using HDC/VSA in the context of emerging computing hardware. 
This section sets up the context for the article. 
Section~\ref{sect:methods} offers a deep dive into the fundamentals of HDC/VSA, recommended primarily to readers not yet familiar with the framework.
Section~\ref{sect:comp:VSAs} explains different aspects of computing with HDC/VSA, including a ``cookbook'' for the representation primitives for numerous data structures (Section~\ref{sect:primitives}) as well as introducing an idea of computing in superposition and its existing applications (Section~\ref{sect:comp:super}). 
Current hardware realizations of HDC/VSA models are considered in Section~\ref{sect:hardware}.
Section~\ref{sect:related} provides the discussion.
Finally, Appendix~\ref{sect:turing} describes proposals for implementing two Turing complete systems with HDC/VSA.

\section{Motivation}
\label{sect:motiv}

The exponential growth of Big Data and AI applications exposes fundamental limitations of the conventional computing framework. One problem is that energy efficiency is stagnating \cite{power_efficiency} -- training and fine-tuning a neural network for a Natural Language Processing application consumes energy and computational resources equivalent to several hundred thousand US dollars~\cite{StrubellEnergyNLP} or more~\cite{RogersNLP}.
Conventional computing hardware is also highly susceptible to errors and energy is often ``wasted'' attempting to maintain low error rates. 

Data-intensive applications illustrate the scale of the problem and make 
energy efficiency the grand challenge of computer engineering.
To solve this challenge, alternative hardware is required that can work with imprecise and unreliable computational elements~\cite{JaegerComputing2020}. 
Operating at ultra-low voltages with stochastic devices that are prone to errors has the potential to greatly increase computing power and efficiency.
For example, the recent advances in materials science as well as in device manufacturing make it possible to design computing hardware that accommodates computational principles of biological brains or exploits physical properties of the substrate material. 
For certain classes of problems, computing hardware such as neuromorphic processors~\cite{TrueNorth14, Loihi18, NNkNN20} and in-memory computing architectures~\cite{MemristorHD19} consumes only a fraction of the energy compared to current technology.
For certain tasks, existing neuromorphic platforms can be $1,000$ times more energy efficient~\cite{Loihi18} than the conventional ones.

There is currently a  focus on implementing AI capabilities in emerging computing  hardware~\cite{NNkNN20}, with the aim of providing an energy-efficient implementation of a selected class of AI functionalities (mainly neural networks).
However, we see the opportunity for a computational framework exceeding neural networks in scope, which could empower an unprecedented breakthrough in emerging computing technology. 
First, while neural network algorithms serve a rather small subset of computation problems extremely well, they are unable to address a large class of problems that require conventional algorithms and data structures. A computing framework with a broader application scope than neural networks could boost the adoption of emerging computing by several orders of magnitude.  
Second, despite many promising applications for emerging computing hardware, the programming of any new functionality is far from trivial. Emerging computing hardware currently lacks a holistic software architecture, which would streamline the development of the new functionality. 
Current development strategies resemble those of assembly programming, where the developer is left with the entire job -- from coming up with the algorithmic idea to designing the actual machine instructions to be executed by a central processing unit. Thus, the impressive recent emerging hardware development~\cite{MemristorHD19, LiHD16} needs to be complemented with the creation of computing frameworks for such hardware, which can abstract and simplify the implementation of new functionalities, including the design of programs.
Last but not least, most emerging hardware differs fundamentally from traditional computer and neural network accelerator hardware in that the enabled computations are unreliable and stochastic. Thus, a computing framework is required in which error correction and error robustness are achieved.

There is ample work demonstrating that HDC/VSA possesses a rich computational expressiveness, from the functionality of neural networks~\cite{intSOM, intESN, intRVFL,Frady17} to machine learning tasks~\cite{RPRSKJ2015, Rasanen2015tr, TNNLS18, HDGestureIEEE,Rachkovskij2022NCA} and cognitive modeling~\cite{RachkovskijAnalogical2004, RachkovskijWorldModel2013, BuildBrain, RachkovskijSimilarity2012, EmruliAnalogical2013, KleykoBees2015}.
Further, HDC/VSA can express conventional algorithms, for example, finite state automata~\cite{HD_FSA,UCBHD_FSA} and context-free grammars~\cite{VSAGrammars}.

In this article, we explore whether HDC/VSA can serve as a computing framework for taking emerging computing to the next level.  
We argue that HDC/VSA provide a framework to formalize and modularize algorithms and, at the same time, bridge the computation and implementation levels in Marr's framework~\cite{MarrVision1982} for information processing systems  (see Fig.~\ref{fig:stack}). Our proposal generalizes earlier suggestions to apply HDC/VSA for implementing specific machine learning algorithms on emerging hardware~\cite{HDNP17, Kanerva19}.
 
\section{Fundamentals of HDC/VSA}
\label{sect:methods}

HDC/VSA~\cite{Gayler2003,Kanerva09}, 
is the term for a family of models for representing and manipulating data in a high-dimensional space. 
It was originally proposed in cognitive psychology and cognitive neuroscience as a connectionist model for symbolic reasoning~\cite{PlateAnalogical1994}.
In HDC/VSA, data objects are represented by vectors of high (but fixed) dimension $N$, sometimes called hypervectors or HD vectors. The encoded information is distributed across all components of a hypervector. 
Such distributed representations~\cite{Hinton1986} are distinct from localist and semi-localist representations~\cite{ThorpeLocalized1998}, where single or subsets of components encode individual data objects.

Distributed representations are, in and of themselves, not the full story.  
As argued by \cite{FodorCritical1988}, distributed representations must be 
productive and systematic.
Productivity refers to massive expressiveness generated by simple primitives, while systematicity means that representations are sensitive to the structure of the encoded objects.
These desiderata were one of the drivers for developing HDC/VSA.
One major advantage of HDC/VSA as the algorithmic level in the Marr hierarchy  (Fig.~\ref{fig:stack}) is that it embraces distributed representations, which are robust to local noise.

The idea of computing with random hypervectors as basic objects rather than Boolean or numeric scalars was developed by Kussul as part of Associative-Projective Neural Networks~\cite{RachkovskijTexture1991} and independently in seminal works by Smolensky on Tensor Product Variable Binding~\cite{Smolensky1990} \& Plate on Holographic Reduced Representation~\cite{PlateThesis}. 
HDC/VSA can be formulated with different types of vectors, namely those containing real, complex, or binary entries, as well as with the multivectors from geometric algebra.
These HDC/VSA models come under many different names: Holographic Reduced Representation (HRR)~\cite{PlateTr,PlateBook}, 
Multiply-Add-Permute (MAP)~\cite{MAP}, 
Binary Spatter Codes~\cite{Kanerva97},
Sparse Binary Distributed Representations (SBDR)~\cite{CDT2001,KleykoSDR2016},
Sparse Block-Codes~\cite{Laiho2015, FradySDR2020},
Matrix Binding of Additive Terms (MBAT)~\cite{Gallant13}, Geometric Analogue of Holographic Reduced Representation (GAHRR)~\cite{GAHRR}, etc.
All of these different models have similar computational properties -- see~\cite{Frady17} and ~\cite{SchlegelVSAComparison2020}. 
For clarity, we will use the Multiply-Add-Permute model in the remainder of this article. 

\subsection{Basic elements of HDC/VSA}
\label{sect:methods:basic}

\subsubsection{High-dimensional space}

HDC/VSA requires a high-dimensional space.
The appropriate choice of dimensionality $N$ is somewhat dependent on the problem, but there are simple rules of thumb ($N > 1,000$, for example), and the representation of particular data structures in the given problem is much more important.
As mentioned above, there are HDC/VSA models defined for different types of spaces (see Section~\ref{sect:frameworks} for more details).
In this article, we will use a variation of the Multiply-Add-Permute model (MAP-I, see, e.g.,~\cite{SchlegelVSAComparison2020})  that operates in integer vector spaces ($\mathbb{Z}^N$).
Operations and properties that have proven useful are presented below (Appendix~\ref{sect:VSA:summary} provides the summary). 
It is worth pointing out that the superposition and binding of
hypervectors form an algebraic structure that resembles a field, and
that permutations extend the algebra to all finite groups up to size
$N$.

\subsubsection{Quasi-orthogonality}
\label{sec:vasic:orthogonality}

HDC/VSA uses random (strictly speaking, pseudo-random) vectors as a means for data representation.
By using random vectors as representations, HDC/VSA can exploit the concentration of measure phenomenon \cite{ledoux2001concentration,Gorban2018Blessing}, which implies that with high probability random vectors become almost orthogonal in high-dimensional vector spaces. This phenomenon is sometimes called progressive precision~\cite{ComputingRandomness} or the blessing of dimensionality~\cite{Gorban2018Blessing}.
In the case of HDC/VSA, it means that when, e.g., two objects are represented by random vectors, with high probability their representations will be almost orthogonal to each other.
Multiply-Add-Permute uses bipolar random vectors where the $i$-th component of a vector $\mathbf{a}$ is generated i.i.d. random from the Bernoulli distribution: $a_i \sim 2\mathcal{B}(0.5)-1$.
In the HDC/VSA literature, dissimilar representations are described by various adjectives such as unrelated, uncorrelated, approximately-, pseudo-, or quasi-orthogonal.
Unlike exact orthogonality, the dimension $N$ is not a hard limit on the number of quasi-orthogonal vectors one can create.

\subsubsection{Similarity measure}
Processing in HDC/VSA is based on similarity between hypervectors. 
The common similarity measures in HDC/VSA are the dot (scalar, inner) product, cosine similarity, overlap, and Hamming distance. 
In Multiply-Add-Permute, it is common to use either the cosine similarity or the dot product. 
Therefore, we will be using the dot product (denoted as $\langle\cdot,\cdot\rangle$)
as the similarity measure below.

\subsubsection{Seed hypervectors}

When designing an HDC/VSA algorithm for solving a problem, it is common to define a set of the most basic concepts/symbols for the given problem and assign hypervectors to them. 
Such seed hypervectors are defined as the representations of concepts that are irreducible. All other hypervectors occurring in the course of a computation are therefore reducible, that is, they are composed of seed hypervectors.
Here we will focus on symbolic structures, i.e., symbols from some alphabet with size $D$, which are represented by i.i.d. random seed hypervectors (see Section~\ref{sec:vasic:orthogonality}). 
As mentioned above, in Multiply-Add-Permute, seed hypervectors are bipolar and so any hypervector $\mathbf{a} \in  \{-1,1\}^N$.
The process of assigning seed hypervectors, usually (but not always) by i.i.d. random generation of vectors, is referred to as mapping, encoding, projection, or embedding. 
We reiterate that representations in an HDC/VSA algorithm need not always be quasi-orthogonal.
For example, for representing real-valued variables one might use a locality-preserving representation scheme, in which representations of similar values are systematically correlated and not quasi-orthogonal~\cite{Scalarencoding, WeissOlshausenSpatial16, KomerContinuous2019},
or where the hypervectors are learned~\cite{RPRSKJ2015, Sutor18}. 
Thus, one should keep in mind that i.i.d. randomness is not the only tool for designing seed representations.

\subsubsection{Item memory}

Seed hypervectors are stored in the so-called item memory (or cleanup memory), a content-addressable memory which can be just a matrix or an associative memory~\cite{FrolovWillshaw2002, FrolovTime2006, SurveyAM17} that
stores the hypervectors as point attractors.

\subsection{HDC/VSA operations and compound representations}
\label{sect:oper}

Seed hypervectors are the building blocks for compound HDC/VSA representations, which are built from operations performed on the seed vectors.
For example, a compound hypervector representing edges of a graph (compound entity) can be constructed (Section~\ref{sect:graphs}) from seed hypervectors representing its nodes (basis symbols).
This compositional formation of data structures in HDC/VSA is akin to conventional computing and very different from the modern neural networks in which activity vectors, especially in hidden layers, often can not be readily parsed.

Two key HDC/VSA operations are dyadic vector operations between hypervectors that are referred to as {\em superposition} and {\em binding}. Like the corresponding operations between ordinary numbers, they form, together with the representation vector space, a field-like algebraic structure. Another important HDC/VSA operation is the {\em permutation} of components within a hypervector.

The component-wise addition operation is used for bundling or superposing and in the Multiply-Add-Permute model it is implemented as a component-wise addition of hypervectors.
The binding operation is used for
variable binding. In the Multiply-Add-Permute model, the binding operation is implemented via component-wise multiplication, i.e., via the Hadamard product. 
The permutation operation, as its name suggests, shuffles the components of a hypervector according to a pre-defined permutation that can be, e.g., chosen randomly.  
In practice, a rotation of components, i.e., a cyclic shift of the hypervector component index, is used frequently. 

In what follows, we describe each operation and its properties in more detail.
It is important to stress that various HDC/VSA models differ in the particular details of realizing their operations.
As a consequence, the operations' properties  presented below are relevant for the Multiply-Add-Permute model but are not valid for each and every HDC/VSA model.
For the sake of focus, we will not discuss differences between different HDC/VSA models in depth here, but we encourage interested readers to consult recent studies~\cite{SchlegelVSAComparison2020, KleykoSurveyVSA2021Part1}.

Note also that the seed hypervectors referred to in this section are pseudo-random i.i.d.  
Because high-dimensional representation tolerates errors, the conditions
listed below need only be satisfied approximately or with high
probability.
Due to the concentration of measure phenomenon, the operations -- and computations based on them -- become ever more reliable, dependable, and predictable as
the dimensionality $N$ of the space increases.

\subsubsection[nocolon]{Binding}  a dyadic operation mapping two hypervectors to another hypervector. It is used to represent an object formed by the binding of two other objects.
This operation is an important ingredient for forming compositional structures with distributed representations (see, e.g., a discussion on its importance in the context of deep learning in~\cite{GreffBinding2020}).
Formally, for two objects $a$ and $b$, represented by the hypervectors $\textbf{a}$ and $\textbf{b}$, the hypervector that represents the bound object (denoted by $\textbf{m}$) is: 
\noindent
\begin{equation}
\label{eq:bind} 
\textbf{m} = \textbf{a} \odot \textbf{b}.  
\end{equation}
\noindent
In the Multiply-Add-Permute model, $\odot$ denotes the component-wise multiplication (Hadamard product). Multiple application of binding is denoted by $\prod$, enabling the formation of a hypervector representing the product of more than two hypervectors.

Consider the example of representing a database for trivia about countries~\cite{Kanerva2010}.  The database record for a country contains the name, the capital, and the currency.  
The first step is to form hypervectors that represent key-value pairs, which can be done by binding:
$\mathrm{\bf{country}} \odot \mathrm{\bf{USA}}$, 
$\mathrm{\bf{capital}} \odot \mathrm{\bf{Washington}}$,
$\mathrm{\bf{currency}} \odot \mathrm{\bf{USD}}$.
To create a single hypervector that represents the entire data record for a country, we need another operation to combine the different key-value pairs (see below).

\subsubsection{Superposition}
\label{sec:vsas:add}
 a dyadic operation mapping two hypervectors to another hypervector.
It is denoted with $+$ and, in the Multiply-Add-Permute model, implemented via component-wise addition, which sometimes can be thresholded to keep bipolar representations (not used in this article). 
The superposition operation combines several hypervectors into a single hypervector. 
For example, for $\textbf{a}$ and $\textbf{b}$  the result $\textbf{z}$ of the superposition of their hypervectors is simply: 
\noindent
\begin{equation}
\label{eq:bindle} 
\textbf{z} = \textbf{a} + \textbf{b}.
\end{equation}
The superposition of more than two hypervectors is denoted by $\sum$.
\noindent
Often, superposition is followed by a thresholding operation to produce a resultant hypervector that is of the same type as the seed vectors.
For example, in the Multiply-Add-Permute model the seed hypervectors are bipolar vectors, but the arithmetic sum-vector is not. 
Therefore, in the bipolar variant (MAP-B, see~\cite{SchlegelVSAComparison2020}) a thresholding operation, using the signs in each component, can map the sum vector back to a bipolar hypervector. 
This type of thresholding is sometimes called the majority rule/sum and denoted by brackets: $[\textbf{a} + \textbf{b}]$. 
For the sake of consistency, the examples below use the non-thresholded sum, unless mentioned otherwise.

The non-thresholded sum has the advantage of being invertible since individual elements in the sum can be removed by subtraction (denoted as $-$) without interfering with the rest. 
Using the example above: 
\noindent
\begin{equation}
\label{eq:subtract} 
\textbf{a} = \textbf{z}  - \textbf{b}.
\end{equation}

Continuing the database example, the superposition operation can be used to create a single hypervector from hypervectors representing all key-value pairs of the record. 
Thus, the compound hypervector for the whole record will be formed as: $\mathrm{\bf{country}} \odot \mathrm{\bf{USA}} + 
\mathrm{\bf{capital}} \odot \mathrm{\bf{Washington}} + \mathrm{\bf{currency}} \odot \mathrm{\bf{USD}}$.

\subsubsection{Permutation} 
a unary operation on a hypervector that yields a hypervector.
Akin to the binding operation, permutation is often used to map into an area of hypervector space that does not interfere strongly with other representations. 
However, unlike binding in Multiply-Add-Permute, the same permutation can be used recursively, projecting into previously unoccupied space with every iteration.   
Note that the number of possible permutations grows super-exponentially with the dimensionality ($N!$) and that permutations themselves are not elements of the space of representations. 
In most HDC/VSA algorithms, a single one or a small set of permutations are fixed at the onset of computation. We continue with a simple example, and more examples follow in the subsequent sections.

Permutation can be seen as an alternative approach to binding when there is only one hypervector as the operand~\cite{MAP}.
The permutation operation can also be used to represent sequence relations and other asymmetric relations like ``part-of''.
For example, a fixed permutation (denoted as $\rho(\cdot)$) can be used to associate, e.g., a symbol hypervector with the position of a symbol in a sequence, resulting in a hypervector representing the symbol in that position.
The single application of the permutation is:
\noindent
\begin{equation}
\label{eq:perm} 
\textbf{r} = \rho^1( \textbf{a} )= \rho( \textbf{a} ).
\end{equation}
\noindent
To associate $\textbf{a}$ with the $i$-th position in a sequence, the permutation is applied $i$ times. The result is the hypervector:
\noindent
\begin{equation*}
\textbf{r} = \rho^i( \textbf{a}).
\end{equation*}
\noindent
Note that permutation is an example of a more general unary operation, matrix-vector multiplications (see, e.g.,~\cite{Gallant13} for a proposal on using matrix-vector multiplications to implement the binding operation).

\subsubsection{Properties of HDC/VSA operations and their interaction}
\label{sec:VSA:properties}
Here we summarize the properties of the basic HDC/VSA operations and how they interact:

\paragraph{Superposition}
\begin{itemize}
    \item  Superposition can be inverted with subtraction: 
    $\textbf{a} + \textbf{b} + \textbf{c} - \textbf{c} =  \textbf{a} + \textbf{b}$;    
    \item  In contrast to the binding and permutation operations, the result of the superposition  $\textbf{z}=\textbf{a}+\textbf{b}$ (often called the superposition hypervector) is similar to each of its argument hypervectors, i.e., the dot product between $\textbf{z}$ and $\textbf{a}$ or $\textbf{b}$ is significantly more than $0$: 
    $  \langle \textbf{z}, \textbf{a} \rangle \approx \langle \textbf{z}, \textbf{b} \rangle >  0$;
    \item  Arguments of binding can be approximately recovered from the superposition hypervector: 
    $\textbf{b} \odot 
    (\textbf{a} \odot \textbf{b} + \textbf{c} \odot \textbf{d}) \approx  \textbf{a}$;
    \item Superposition is commutative: $\textbf{a} + \textbf{b} = \textbf{b} + \textbf{a}$;
    \item Thresholded superposition is approximately associative: 
    $[[\textbf{a} + \textbf{b}] + \textbf{c}] \approx [\textbf{a} + [ \textbf{b} + \textbf{c}]] $.
\end{itemize}
Note that if several instances of any hypervector are included (e.g., $\textbf{z} = 3\textbf{a} + \textbf{b}$), the resultant hypervector is more similar to the dominating hypervector than to other arguments.

\paragraph{Binding}
\label{sect:mulp:prop}

\begin{itemize}
    \item Binding is commutative: $\textbf{a} \odot \textbf{b} = \textbf{b} \odot \textbf{a}$;
    \item Binding distributes over superposition: $\textbf{c} \odot (\textbf{a} + \textbf{b}) = (\textbf{c} \odot \textbf{a}) + (\textbf{c} \odot \textbf{b})$;
    \item Binding is invertible, for $\textbf{m} =  \textbf{a} \odot \textbf{b}$: $ \textbf{a} \odot \textbf{m} = \textbf{b}$. The inversion process is often called releasing or unbinding. In the case of the  component-wise multiplication of bipolar vectors, the unbinding operation is performed with the same operation.
    Therefore, we do not introduce a separate notation for unbinding here;
    \item Binding is associative: $\textbf{c} \odot (\textbf{a} \odot \textbf{b}) = (\textbf{c} \odot \textbf{a}) \odot \textbf{b}$;
    \item The result of binding is dissimilar to each of its argument hypervectors, e.g., $\textbf{m}$ is dissimilar to the hypervectors being bound, i.e., the dot product between $\textbf{m}$ and $\textbf{a}$ or $\textbf{b}$ is approximately $0$: $ \langle\textbf{m}, \textbf{a} \rangle \approx \langle \textbf{m}, \textbf{b} \rangle \approx  0$;
    \item Binding preserves similarity (for similar $\textbf{a}$ and $\textbf{a}'$): 
    $\langle \textbf{a} \odot \textbf{b}, \textbf{a}' \odot \textbf{b}  \rangle \gg 0$;
    \item Binding is a ``randomizing'' operation (since 
    $\langle \textbf{a} \odot \textbf{b}, \textbf{a} \rangle \approx 0$) that preserves similarity (because $ \langle \textbf{a} \odot \textbf{b}, \textbf{c} \odot \textbf{b} \rangle = \langle \textbf{a}, \textbf{c} \rangle $).
\end{itemize}

\paragraph{Permutation}
\begin{itemize}
    \item  Permutation is invertible, for $\textbf{r} = \rho( \textbf{a} )$: $\textbf{a} = \rho^{-1}(\textbf{r})$;
    \item  In Multiply-Add-Permute, permutation distributes over both binding ($\rho (\textbf{a} \odot \textbf{b}) = \rho ( \textbf{a}) \odot \rho ( \textbf{b})$) and superposition ($\rho (\textbf{a} + \textbf{b}) = \rho ( \textbf{a}) + \rho ( \textbf{b})$);
    \item Similar to the binding operation, the result $\textbf{r}$ of a (random) permutation  is dissimilar to the argument hypervector $\textbf{a}$: 
    $ \langle \textbf{r}, \textbf{a} \rangle \approx 0$;
    \item Permutation is a  ``randomizing'' operation (since 
    $ \langle \rho( \textbf{a}), \textbf{a} \rangle \approx 0$) 
    that preserves similarity (because 
    $ \langle \rho( \textbf{a}), \rho( \textbf{b}) \rangle = \langle \textbf{a}, \textbf{b} \rangle $);
\end{itemize}

It is worth clarifying what we mean by ``similarity preserving'' in the case of binding and permutation vs. superposition above:  For binding, the similarity between two hypervectors is the same
before and after binding with a third hypervector, i.e.,
$ \langle \textbf{a} \odot \textbf{b}, \textbf{c} \odot \textbf{b} \rangle = \langle \textbf{a}, \textbf{c} \rangle $, and for permutation,
the similarity between two hypervectors is also the same
before and after the operation, i.e., 
$\langle \rho( \textbf{a}), \rho( \textbf{b})  \rangle = 
 \langle \textbf{a}, \textbf{b}  \rangle $.
However, for superposition, the similarity between two hypervectors is generally lower before vs. after superimposing them to a third hypervector, i.e., 
$\langle \textbf{a} + \textbf{b}, \textbf{c} + \textbf{b} \rangle > \langle\textbf{a}, \textbf{c} \rangle$, 
since the sum moves them in a common direction $\textbf{b}$.  
On the other hand, since the superposition hypervector is similar to each of the vectors in the sum,
$ \langle \textbf{a}+\textbf{b}, \textbf{a} \rangle \approx 
\langle \textbf{a}+\textbf{b}, \textbf{b} \rangle >  0$, 
it is also sometimes referred to as ``similarity preserving,'' in contrast to binding and permutation, which generally create a dissimilar hypervector. One should keep this distinction in mind when referring to the similarity preserving properties of these operators.

\subsection{
Parsing compound representations}
\label{sect:probing}

HDC/VSA offer the possibility to encode data structures into compound hypervectors and to manipulate the hypervectors with the operations described above to perform computation on the data structures. 
In conventional computing, data structures are always exposed and the algorithm queries or modifies individual elements within them. In contrast, the vector operations in HDC/VSA can search or transform  many or all elements of a data structure in parallel, which we call ``{\it computing in superposition}'' (see Section~\ref{sect:comp:super}). 
All data structures are hypervectors and can be manipulated immediately and in parallel, regardless of how complicated a structure they possess. 
But this also means that the data structure of a compound hypervector is not immediately decodable from the item memory. 
To query element(s) of a compound hypervector, it first needs to be analyzed or ``parsed''. We borrow the term parsing from linguistics because the parsing of HDC/VSA hypervectors is somewhat similar. To understand a sentence, one needs to divide the sentence into its parts and assign their syntactic roles, which involves comparing the parts with the stored information about their meaning and syntactic roles. 
Similarly, to extract the result of a HDC/VSA computation, one has to parse the resultant hypervector. The parsing of HDC/VSA hypervectors involves the decomposition and comparison of the resulting parts with the stored information.

Like with the sum or product of ordinary numbers, the parsing of hypervectors requires additional information, such as the operations used to form the compound representation and the set of seed vectors.  
Parsing a compound hypervector first entails the operation inverse that used to encode the wanted element in the data structure. However, the result is almost always approximate because of crosstalk noise coming from all the other elements in the compound hypervector.
To single out the correct result, the noisy vector has to be compared to the original seed vectors in terms of similarity. 
Probing is the process of retrieving the best-matching hypervector (i.e., the nearest neighbor) among the hypervectors for a given query hypervector.  This is done in the item memory, which contains all the seed hypervectors. 
For example, consider the compound hypervector: 
\noindent
\begin{equation*}
\textbf{s} = \textbf{a} \odot \textbf{b} +  \textbf{c} \odot \textbf{d} .
\end{equation*}
\noindent
In order to know which hypervector has been bound to, e.g., $\textbf{b}$ we have to unbind (inverse binding) $\textbf{b}$ from $\textbf{s}$:
\noindent
\begin{equation*}
\begin{split}
\textbf{s} \odot \textbf{b} & =  \textbf{b}  \odot (\textbf{a} \odot \textbf{b} +  \textbf{c} \odot \textbf{d} )= \\
&=\textbf{a} + 
\textbf{b}  \odot \textbf{c} \odot \textbf{d} =  \textbf{a}   + \mathrm{noise} \approx \textbf{a}.
\end{split}
\end{equation*}
\noindent
The resultant hypervector contains the correct answer $\textbf{a}$ and a crosstalk noise term $\textbf{b}  \odot \textbf{c} \odot \textbf{d}$, which is dissimilar to any of the items in the item memory.
The query hypervector $\textbf{a}   + \mathrm{noise} $ will be highly similar to the hypervector $\textbf{a}$ stored in the item memory, which will be successfully retrieved by the nearest neighbor search with high probability.
Thus, the probing operation removes (or cleans up) the noise and returns the correct result.

Cleanup via probing is a critical part of HDC/VSA computations and has the advantage that its operation is intrinsically noise resilient and the degree of noise robustness can be easily controlled by the dimension $N$. 
In essence, probing is a signal detection problem. 
The number of hypervectors that can be correctly retrieved from the superposition is called capacity. The capacity increases roughly linearly with the hypervector dimension and is quite insensitive to the details of a particular HDC/VSA model. The signal detection theory for HDC/VSA~\cite{Frady17} enables one to determine the dimension of the hypervector space that is sufficient for a given computation and a given precision of the hardware.

\subsubsection*{Parsing hypervectors with multiple bindings}
\label{sec:resonator:network}
In the example above, it was assumed that one argument (i.e., $\mathbf{b}$) of the key-value pair  was known. 
This, however, is not always the case. 
Moreover, there exist representations where several hypervectors are being bound (e.g., $\mathbf{a} \odot \mathbf{b} \odot \mathbf{c}$).  
Parsing compound hypervectors with such elements is challenging due to the fact that the binding operation in the Multiply-Add-Permute model
produces a hypervector dissimilar to its arguments (cf. Section~\ref{sect:mulp:prop}). 
This means that the most obvious way to parse hypervectors of the form $\mathbf{a} \odot \mathbf{b} \odot \mathbf{c}$ is by brute force, through checking all possible combinations of the arguments. 
The number of such combinations, however, grows exponentially with the number of arguments involved.
Therefore, a mechanism called a resonator network has been proposed~\cite{ResPart1, ResPart2}, which addresses this problem by a parallel search in the space of all possible combinations.

\begin{figure}[tb]
\centering
\includegraphics[width=1.0\columnwidth]{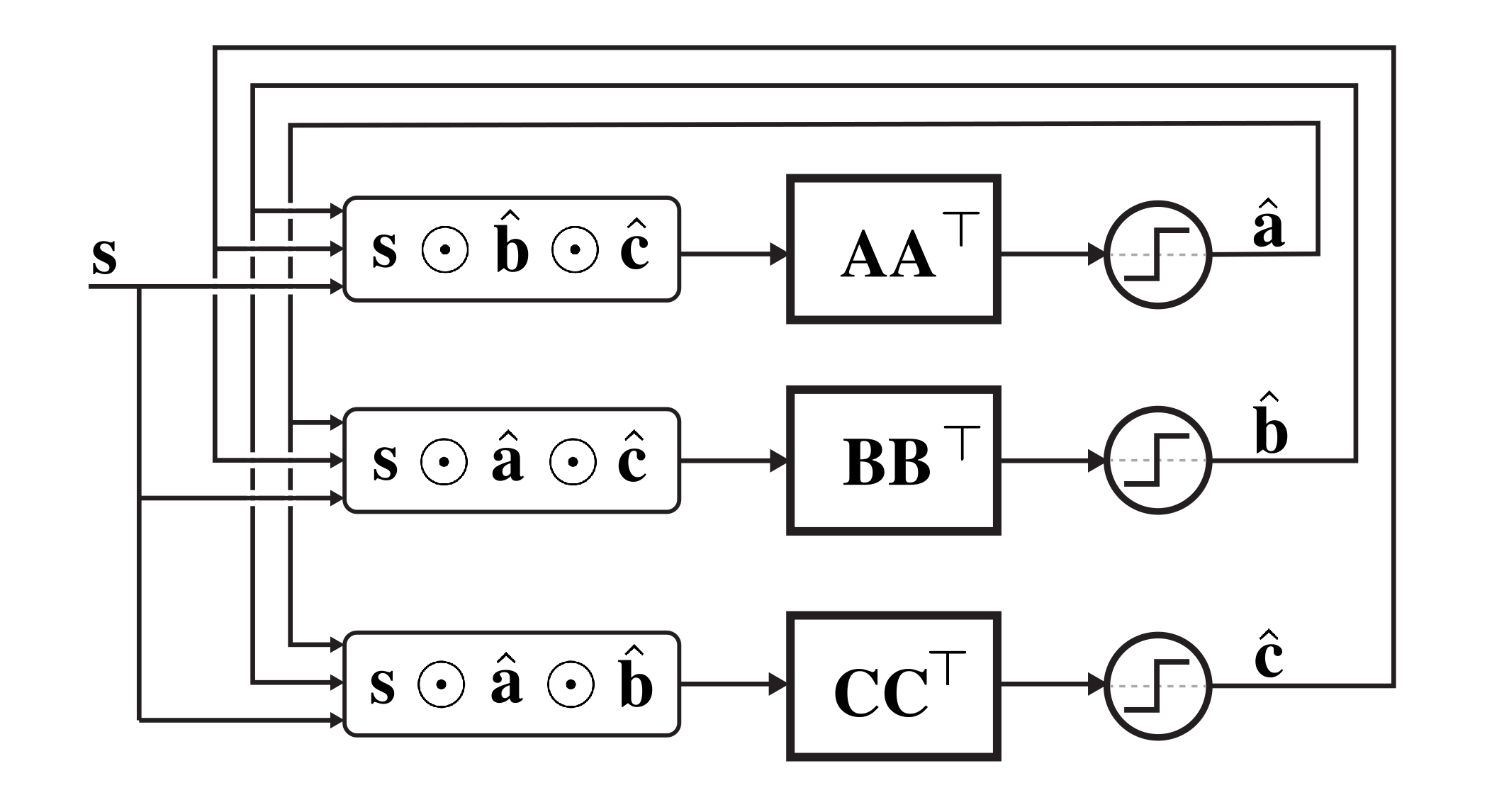}
\caption{
An example of a resonator network with three arguments. It is factoring a compound hypervector $\mathbf{s} = \mathbf{a} \odot \mathbf{b} \odot \mathbf{c}$;
$\mathbf{A}$, $\mathbf{B}$, and $\mathbf{C}$ denote the corresponding item memories containing seed hypervectors for $\mathbf{a}$, $\mathbf{b}$, and $\mathbf{c}$ arguments, respectively. 
}
\label{fig:resonator}
\end{figure}

The resonator network assumes that none of the arguments are given, but that they are contained in different item memories, which should be known to the resonator network. 
Fig.~\ref{fig:resonator} illustrates an example of a resonator network for factoring the hypervector $\mathbf{s} = \mathbf{a} \odot \mathbf{b} \odot \mathbf{c}$.
In a nutshell, the resonator network is a novel recurrent neural network design that uses HDC/VSA principles to solve combinatorial optimization problems. 
As shown in the example, it factors the arguments of the input vector $\mathbf{s}$ representing the binding of several hypervectors. 
To do so it uses hypervectors $\mathbf{\hat{a}}(t), \mathbf{\hat{b}}(t), \mathbf{\hat{c}}(t),$ each storing the prediction for a particular argument of the product forming $\mathbf{s}$. 
Each prediction communicates with the input hypervector ($\mathbf{s}$) and all other predictions using the following dynamics:
\noindent
\begin{equation}
\begin{split}
&\mathbf{\hat{a}}(t+1)= \text{sign} \Big( \mathbf{A} {\mathbf{A}}^{\top} (\textbf{s} \odot  \mathbf{\hat{b}}(t)  \odot  \mathbf{\hat{c}}(t) ) \Big); \\
&\mathbf{\hat{b}}(t+1)= \text{sign} \Big( \mathbf{B} {\mathbf{B}}^{\top} (\textbf{s} \odot  \mathbf{\hat{a}}(t)  \odot  \mathbf{\hat{c}}(t) ) \Big); \\
&\mathbf{\hat{c}}(t+1)= \text{sign} \Big( \mathbf{C} {\mathbf{C}}^{\top} (\textbf{s} \odot  \mathbf{\hat{a}}(t)  \odot  \mathbf{\hat{b}}(t) ) \Big), 
\end{split}
\label{eq:resnet:text}
\end{equation}
\noindent
where $\mathbf{A}$, $\mathbf{B}$, and $\mathbf{C}$ denote the corresponding item memories containing $\mathbf{a}$, $\mathbf{b}$, and $\mathbf{c}$ arguments, respectively, and $\text{sign}(\cdot)$ denotes a step that projects the predictions back to the bipolar values.
Note that the resonator network does not have to work with only bipolar hypervectors. 
Rather, the usage of the $\text{sign}(\cdot)$ function is determined by the fact that the seed hypervectors in the Multiply-Add-Permute model are bipolar.
Thus, other types of nonlinearity functions can be used to make a resonator network compatible with the desirable format of the seed hypervectors. 
Note also that these item memories will contain other hypervectors as well, but hypervectors stored in $\mathbf{A}$, $\mathbf{B}$, and $\mathbf{C}$ differ from each other. 
The size of each item memory depends on a task but it will affect the performance of the resonator network as larger item memories imply a larger search space. 

The key insight into the internals of the resonator network is that it iteratively tries to improve its current predictions of the arguments constituting the input hypervector $\mathbf{s}$.  
In essence, at time $t$ each prediction might hold multiple weighted guesses from  the corresponding item memory. 
The current predictions for other arguments are used to invert the input vector and infer the current argument (e.g, $\textbf{s} \odot  \mathbf{\hat{b}}(t)  \odot  \mathbf{\hat{c}}(t)$).
The cost of using the superposition for storing the predictions is crosstalk noise.
To clean up this noise, the predictions are projected back to their item memories (e.g., ${\mathbf{A}}^{\top} (\textbf{s} \odot  \mathbf{\hat{b}}(t)  \odot  \mathbf{\hat{c}}(t) )$), which provides weights for different seed hypervectors stored in the item memory and, therefore, constrains the predictions to only to the valid entries in the item memory.
These weights are then used to form a new prediction, which is a weighted superposition of all seed hypervectors.  
Successive iterations of the process in Eq.~(\ref{eq:resnet:text}) eliminate the noise as the arguments become identified and find their place in the input vector. 
Once the arguments are fully identified, the resonator network reaches a stable equilibrium and the arguments can be read.
For the sake of space, we do not go into the details of applying resonator networks here. 
Please refer to~\cite{ResPart1} for examples of factoring hypervectors of data structures with resonator network and to~\cite{ResPart2} for their comparison with other standard optimization-based methods.

\subsection{Generality and utility}

Currently, there are several known areas where HDC/VSA have been employed.
Hypervectors serve as representations for cognitive architectures~\cite{RachkovskijWorldModel2013, BuildBrain}.
They are used  for the approximation of conventional data structures~\cite{HD_FSA, UCBHD_FSA,  ABF}, distributed systems~\cite{SimpkinHDWorkflow2019, RosatoHDDistributed2021}, communications~\cite{JakimovskiCollective2012, KleykoMACOM2012, KimHDM2018}, for forming representations in natural language processing applications~\cite{RPRSKJ2015, RIJHK2015} and robotics~\cite{AAAIWLevyBajracharyaGaylor, Neubert2019, Mitrokhin19HD, HAPCOM20, HerscheDVSCDT2020}.
The fact that it is possible to map real-valued data to hypervectors allows one to apply HDC/VSA in machine learning domains.
Most of these works were connected to classification tasks (see a recent overview in~\cite{HDClassGe}).
Examples of domains that have benefited from the application of HDC/VSA modeling are biomedical signal processing~\cite{HDGestureIEEE, ACCESS_HRV}, gesture recognition~\cite{HD_ICRC16, TNNLS18}, seizure onset detection and localization~\cite{BurrelloLBPSeizure2020}, physical activity recognition~\cite{Rasanen14}, and fault isolation~\cite{ACCESS_BIOFAULT}. 
However, HDC/VSA modeling can also be useful for very generic classification tasks~\cite{intRVFL, DiaoGLVQHD2021}.
The common feature of these works is a simple training process, which does not require the use of iterative optimization methods, and transparently learns with few training examples.

\section{Computing with HDC/VSA}
\label{sect:comp:VSAs}

\subsection{Computational primitives formalized in HDC/VSA}
\label{sect:primitives}

In the previous section, we have introduced the basic elements of HDC/VSA. 
To provide the algorithmic level in the Marr computing hierarchy in Fig.~\ref{fig:stack}, one needs to combine elements of HDC/VSA into primitives of HDC/VSA computing, i.e., something akin to design patterns in software engineering.
For instance, a set of HDC/VSA templates has been proposed for tasks in the domain of personalized devices covering different multivariate modalities such as electromyography, electroencephalography, or electrocorticography~\cite{HDGestureIEEE}.
Here we summarize best practices for representing well-known data structures with HDC/VSA -- this section can be thought of as a ``HDC/VSA cookbook''. 
All examples in this section are available in an interactive Jupyter Notebook\footnote{
\url{https://github.com/denkle/HDC-VSA_cookbook_tutorial}
}. 
After providing some basic rules for representing data structures with HDC/VSA, we present a collection of primitives from prior work that has been done along these lines.
We do not go into an advanced topic of how distributed representations of data structures can be used to construct or learn single-shot transformation between data structures that share symbols.
It is, however, worth noting that this property differentiates distributed representations from conventional data structure manipulations and the interested readers are referred to, e.g.,~\cite{NeumannTransformation2002, PlateStructure1997} for more details.   
A well-known example of this property has been presented in~\cite{KanervaLearning2000} where a mapping between the ``mother-of'' relation to the ``parent-of'' relation was constructed with simple vector operations and using only a few examples. 
It was shown later in~\cite{KleykoBees2015} that such a mapping can be used to easily form associations between observed structures and decisions caused by these structures.

It is worth noting that in this article we do not cover the representation of real-valued data (see, e.g.,~\cite{KussulRandom1999, Scalarencoding, RachkovskijSimilarityRP2015, Widdows15, FradyFunctions2021}) or solving machine learning problems (see, e.g.,~\cite{HDClassGe}) as it has been covered elsewhere and is outside the immediate scope of the article.

\subsubsection{The rules of thumb}
\label{sec:thumb:rule}
We should point out that the HDC/VSA implementations we describe are not the only possibilities and other solutions may be possible/desirable in a particular design context.
The solutions provided are, however, the most common/obvious choices, based on several ``rules of thumb'':
\begin{itemize}
    \item 
    Superposition is used to combine individual elements of a data structure into a set;

    \item 
    Binding is used to make associations between elements, e.g., key-value pairs;

    \item 
    Permutation is used for tagging data elements to put them into a sequential order, such as in time series; 
    \item 
    Permutation is used for protection from the self-inverse property of the binding operation since the hypervector will not cancel out when bound with its permuted version.

\end{itemize}
We will follow these rules most of the time when forming hypervectors for different data structures.

\subsubsection{Sets}
\label{sect:sets}

A set (denoted as $S$) represents a group of elements, for example, $S=\{a,b,c,d,e\}$.
In order to map a set to a hypervector, two steps are required. 
First, an item memory 
storing random hypervectors for each element of a set is initialized.
We will use bold font in notations of hypervectors (e.g., $\textbf{a}$ for ``\textit{a}'') but a more general notation is via the mapping function $\phi(i) \mapsto \mathbf{i}, i \in S$.
Second, a single hypervector (denoted as $\textbf{s}$) is formed that represents the set as the superposition of hypervectors for the set's elements, e.g., for the set above:
\noindent
\begin{equation*}
\textbf{s} = \textbf{a} + \textbf{b} + \textbf{c} +\textbf{d}+ \textbf{e},
\end{equation*}
\noindent
\noindent
\begin{equation}
\label{eq:set}
\textbf{s} = \sum_{i \in S } \phi(i). 
\end{equation}
\noindent
The hypervector \textbf{s} is a distributed representation of the set $S$.
This mapping preserves the overlap between elements of the sets.
For example, set membership can be tested by calculating the similarity between $\textbf{s}$ and the hypervector corresponding to the element of interest. 
If the similarity score is higher than that expected between two random hypervectors, then most likely the element is present in the set. 
This mapping is very similar to a Bloom Filter~\cite{Bloom1970space} (in particular, to Counting Bloom Filter~\cite{COUNTBF}), which is a  well-known randomized data structure for approximate membership query in a set.
Bloom Filters have been recently shown to be a subclass of HDC/VSA~\cite{ABF}, where the superposition operation is implemented via OR and seed hypervectors are sparse, as in the Sparse Binary Distributed Representations~\cite{CDT2001} model. 
While conceptually representation of sets via distributed representations is a simple idea, it is very influential as it has been applied in myriads of engineering problems (see, e.g., a survey in~\cite{TPBF}).

Note that the limitation of the described mapping of sets is that it does not have a simple and exact way of obtaining distributed representations of the intersection or union of two sets.
The exact results can, obviously, be obtained by first parsing distributed representations of the corresponding sets, reconstructing the symbolic versions, computing the union or intersection in the symbolic domain, and finally forming the distributed representation of the result.
There are, however, simple approximations of the operations the require fewer interactions with the symbolic domain.
Both approximations are obtained by the superposition operation on the corresponding set's hypervectors (e.g., $\mathbf{s}_1$ and $\mathbf{s}_2$):
\noindent
\begin{equation*}
\mathbf{s} = \mathbf{s}_1 + \mathbf{s}_2
\end{equation*}
\noindent
The difference is in the way the parsing of the result in $\mathbf{s}$ is done.
In order to parse the intersection of two sets, only the elements with the largest dot products should be retrieved.
So, if the result of the intersection is stored in $I$, which is initially empty ($I=\emptyset$), then for element $i$ with the corresponding entry $\mathbf{H}_i$ in the item memory:    \noindent
\begin{equation*}
    I= 
\begin{cases}
    I\bigcup \{i\},& \text{if $\mathbf{H}_i \mathbf{s}\geq \Theta_i$ } \\
    I\bigcup \{\emptyset\},              & \text{otherwise}
\end{cases}
\end{equation*}
\noindent
where $\Theta_i$ denotes the corresponding threshold.



To retrieve the union ($U=\emptyset$ at start), the elements with the dot products sufficiently different from the noise level should be considered:
\noindent
\begin{equation*}
    U= 
\begin{cases}
    U\bigcup \{i\},& \text{if $\mathbf{H}_i \mathbf{s}\geq \Theta_n$ } \\
    U\bigcup \{\emptyset\},              & \text{otherwise}
\end{cases}
\end{equation*}
\noindent
where $\Theta_n$ denotes the noise level threshold.
Thus, the subtlety for the intersection is that elements present in both sets will have higher similarity then the ones present in only one of the sets (see Section~\ref{sec:vsas:add}).
This property of the superposition operation is in fact used in the next section for representing multisets.

\subsubsection{Multisets/Histograms/Frequency distributions}
\label{sect:freq}

Let us consider how to form a single hypervector of a multiset or a frequency distribution in the form of counts of the occurrences of various elements in some source.
The mapping is essentially the same as in the case of sets in Section~\ref{sect:sets} with the only difference that a hypervector of an element can be present in the result of the superposition operation several times. 
For example, given  $S=(a,a,a,b,b,c)$, hypervector representing the frequency of elements is formed as: 
\noindent
\begin{equation*}
\begin{split}
\textbf{s} & = \textbf{a} + \textbf{a} + \textbf{a} +\textbf{b} + \textbf{b} + \textbf{c} = \\
&=3\textbf{a} + 2\textbf{b} + \textbf{c}.
\end{split}
\end{equation*}
\noindent
Thus, the number of times a hypervector is present in the superposition determines the frequency of the corresponding element in the sequence.
Using $\textbf{s}$ it is possible to estimate either the frequency of an individual element or compare to the frequency distribution of another sequence.
Both operations require calculating the similarity between $\textbf{s}$ and the corresponding hypervector.

\begin{figure}[tb]
\centering
\includegraphics[width=1.0\columnwidth]{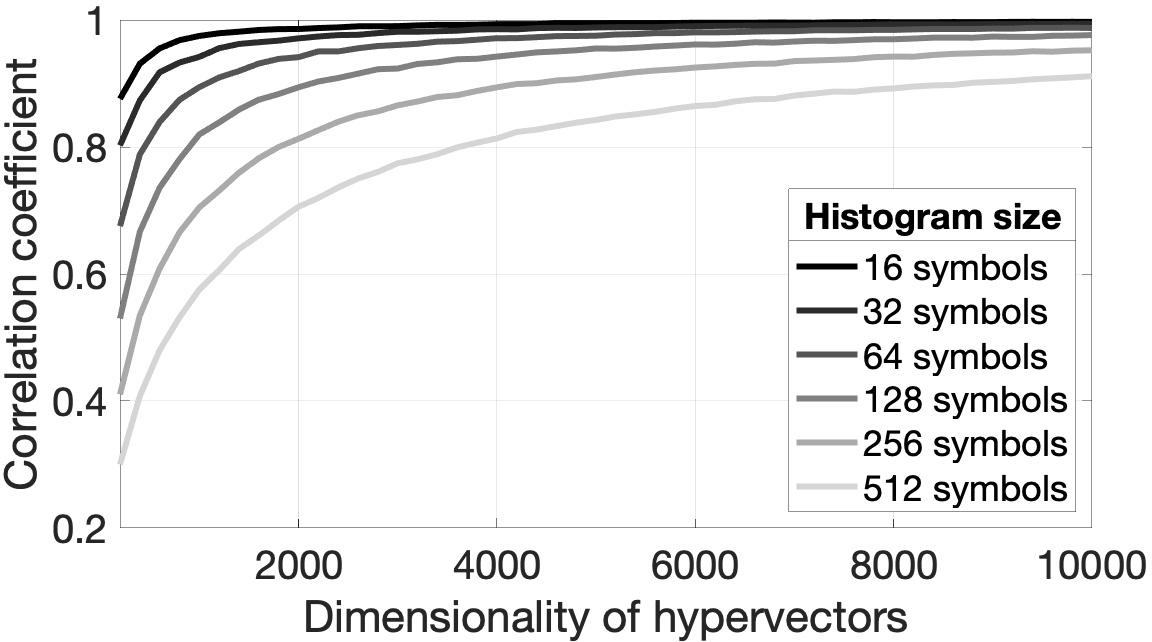}
\caption{
The correlation coefficients between the exact histogram and their approximations from integer-valued $\mathbb{Z}^N$ compound hypervectors.
Six different sizes of histograms were considered. 
The dimensionality of hypervectors varied in the range $[200, 10000]$ with step $200$.
The values of counters were drawn from the discrete uniform distribution $[0, 1023]$.
The reported values were averaged over $100$ simulations.
}
\label{fig:vsas:histograms}
\end{figure}

Usually, $\textbf{s}$ is used as an approximate representation of the exact counters of a histogram.
Fig.~\ref{fig:vsas:histograms} demonstrates Pearson correlation coefficient between the histogram and its approximate version retrieved from a compound hypervector $\textbf{s}$ where the  approximate version was obtained as the dot product between $\textbf{s}$ and symbols' seed hypervectors. 
The simulations were done for different sizes of histogram and varying the dimensionality of hypervectors. 
The results are characteristic for HDC/VSA --  the quality of approximation improved with the increased dimensionality of hypervectors.

This mapping shall be seen as a particular instance of a count-min sketch~\cite{CountMin} that is a randomized data structure for obtaining frequency distributions from sequences.
The count-min sketch is used in a plethora of applications where data are of streaming nature (see, e.g., some examples in~\cite{CountMin}).
Below, in Section~\ref{sect:ngrams} we will also see that the representation of multisets is an essential primitive for representing \textit{n}-gram statistics that in turn is used for solving classification tasks (see, e.g.,~\cite{BurrelloSeizure2019,HyperEmbed, ShridharEnd2End2020}).
The limitation of the presented mapping is that due to the usage of bipolar hypervectors the resultant representation could both overcount and undercount the frequency. 
This limitation is partially addressed by the standard count-min sketch that could only overcount the frequency.

\subsubsection{Cross product of two sets}
\label{sect:tup}

A particularly interesting case is when we have hypervectors representing two different sets (e.g., $\{a,b,c,d,e\}$ and $\{x,y,z\}$). 
Then a mapping based on the binding operation is used to create a hypervector corresponding to the cross product of two sets as follows:
\noindent
\begin{equation*}
\begin{split}
&(\textbf{a} + \textbf{b} + \textbf{c} + \textbf{d} + \textbf{e}) \odot (\textbf{x} + \textbf{y} + \textbf{z} ) = \\
&= (\textbf{a} \odot \textbf{x} + \textbf{a} \odot \textbf{y} + \textbf{a} \odot \textbf{z}) +   
(\textbf{b} \odot \textbf{x} + \textbf{b} \odot \textbf{y} + \textbf{b} \odot \textbf{z}) + \\ 
&+(\textbf{c} \odot \textbf{x} + \textbf{c} \odot \textbf{y} + \textbf{c} \odot \textbf{z}) +  
(\textbf{d} \odot \textbf{x} + \textbf{d} \odot \textbf{y} + \textbf{d} \odot \textbf{z}) + \\
&+(\textbf{e} \odot \textbf{x} + \textbf{e} \odot \textbf{y} + \textbf{e} \odot \textbf{z}).  
\end{split}
\end{equation*}
\noindent

In essence, here occurs (due to the superpositions) a simultaneous binding between all the elements in the two sets.
The cross product set, thus, consists of all possible bindings of hypervectors representing elements of the original sets (e.g., $\textbf{a} \odot \textbf{x}$). 
In the example above, when starting first with the representations of sets, only $7$ operations ($6$ superpositions and $1$ binding) were necessary to form the representation. 
The brute force way for forming the cross product set hypervector would require $29$ operations ($14$ superpositions and $15$ binding).
It is clear that this shortcut works due to the fact that the binding operation distributes over the superposition operation (Section~\ref{sect:mulp:prop}). 
Note that using the Tensor Product Variable Binding~\cite{Smolensky1990} model, the outer product of vector representations of the two sets will be a tensor with the number of dimensions determined by the number of sets in the cross-product.
In contrast, the HDC/VSA representation of a cross-product is given by a hypervector of the same dimension as the individual set hypervectors.
Note also that while it is simple to form a hypervector corresponding to the cross product of two sets with the binding operation, computing the cross product in the symbolic domain might still require lower computational costs as it does not require high-dimensional representations.
Another potential issue of such a representation is the required dimensionality of hypervectors for the situation when all the elements of the cross product should be retrievable from the distributed representation. 
In this case, the dimensionality of hypervectors should be proportional to the product of the sets' cardinalities; so even moderately sized sets require large number of components in hypervectors to provide high accuracy of retrieving individual elements of their cross product from the corresponding hypervector.

\subsubsection{Sequences}
\label{sect:seq}

A sequence is an ordered set of elements. For example, the set from the previous section is now a sequence $(a,b,c,d,e)$, which is not the same as, e.g., $(b,a,c,d,e)$ since the order of elements is different. 
Note that a finite sequence with $k$ elements is called $k$-tuple, with  
an ordered pair being the special case for $k=2$.

Clearly, plain superposition of hypervectors works for representing sets but not 
for sequences, as the sequential order would be lost.
Many authors have proposed the following idea to represent sequences with permutation, e.g., in~\cite{kussul1993information, PlateSequences1995, RPSHK2008, Kanerva09, Frady17, Kanerva19}. Before combining the hypervectors of sequence elements, the order $i$ of each element is associated by applying some specific permutation $k-i$ times to its hypervector (e.g., $\rho^2(\textbf{c})$). 
The advantage of this recursive encoding of sequences is that extending a sequence can be done by permuting $\textbf{s}$ and superimposing or binding it (see below) with the next hypervector in the sequence, hence, incurring a fixed computational cost per symbol. 
The last step is to combine the sequence elements into a single hypervector representing the whole sequence.

There are two common ways to combine sequence elements. 
The first way is to use the superposition operation, similar to the case of sets. 
For the sequence above the resultant hypervector is: 
\noindent
\begin{equation*}
\textbf{s} = \rho^4(\textbf{a}) + \rho^3(\textbf{b}) + \rho^2(\textbf{c}) + \rho^1(\textbf{d}) + \rho^0(\textbf{e}). 
\end{equation*}
\noindent
In general, a given sequence $S$ of length $k$ is represented as: 
\noindent
\begin{equation}
\label{eq:perm_seq}
\textbf{s} = 
\sum_{i=1}^{k}\rho^{k-i}(\phi(S_i)),
\end{equation}
\noindent
where $S_i$ is the $i$th element of sequence $S$.
The advantage of the mapping with the superposition operation is that it is possible to estimate the similarity of two sequences by measuring the similarity of their hypervectors.
Here the similarity of sequences is defined by the number of the same elements in the same sequential positions, where the sequences are aligned by their last elements. 
Evidently, this definition does not take into account the same elements in different positions, in contrast to, e.g., an edit distance of sequences~\cite{levenshtein1966binary}. 
Note that the edit distance can be approximated by vectors of $n$-gram frequencies and their randomized versions akin to hypervectors (see, e.g.,~\cite{Sok2007, HannaganHolographic2011}).

Another advantage of sequence representation with superposition is that it allows easily probing the distributed representation $\textbf{s}$.  
For example, one can check, which element is in position $i$ by applying inverse permutation $i$ times to the resultant hypervector.
Note that permutation of a sequence representation is a general method for shifting an entire sequence by a single operation. 
It produces a shifted sequence where the $i$th element is now at the first position, and thus it can be used to probe the hypervector of element $i$ from the sequence representation. 
For example, when inverting position $3$ in $\textbf{s}$:
\noindent
\begin{equation*}
\begin{split}
 \rho^{-2}(\textbf{s}) & =  \rho^{2}(\textbf{a}) +  \rho^{1}(\textbf{b}) +  \rho^{0}(\textbf{c}) +  \rho^{-1}(\textbf{d}) +\rho^{-2}(\textbf{e})= \\
&= \textbf{c} + \mathrm{noise} \approx \textbf{c}.  
\end{split}
\end{equation*}
\noindent
Probing $\rho^{-2}(\textbf{s})$ with the item memory containing  hypervectors of all sequence elements will return $\textbf{c}$ as the best match (with high probability).

The second way of forming the representation of a sequence involves binding of the permuted hypervectors, e.g., the sequence above is represented as (denoted by $\textbf{p}$): 
\noindent
\begin{equation*}
\textbf{p} = \rho^4(\textbf{a})  \odot \rho^3(\textbf{b})  \odot \rho^2(\textbf{c})  \odot \rho^1(\textbf{d})  \odot \rho^0(\textbf{e}).  
\end{equation*}
\noindent
In general, a given sequence $S$ of length $k$ is represented as: 
\noindent
\begin{equation}
\label{eq:perm_seq1}
\textbf{p} 
=\prod_{i=1}^{k}\rho^{k-i}(\phi(S_i)). 
\end{equation}
\noindent
The advantage of this sequence representation is that it allows forming unique hypervectors even for sequences that differ in only one position.
Section~\ref{sect:ngrams} provides a concrete example of a task where this advantage is important.

Both mappings allow replacement of an element at position $i$ in the sequence if the current element at the $i$th position is known. 
When the superposition operation is used, the replacement requires subtraction of the permuted hypervector of the current element followed by superposition of the permuted hypervector of the new element.
For example, replacing ``d'' to ``z'' in position $4$ is done as follows:
\noindent
\begin{equation*}
\textbf{s} - \rho^1(\textbf{d}) + \rho^1(\textbf{z}) = \rho^4(\textbf{a}) + \rho^3(\textbf{b}) + \rho^2(\textbf{c}) + \rho^1(\textbf{z}) + \rho^0(\textbf{e}).
\end{equation*}
\noindent

When the binding operation is used in the mapping, replacement requires application of the unbinding operation between the permuted hypervector of the current element and $\textbf{s}$, followed by binding with the permuted hypervector of the new element. 
For the example above: 
\noindent
\begin{equation*}
\textbf{s} \odot  \rho^1(\textbf{d})  \odot  \rho^1(\textbf{z}) = \rho^4(\textbf{a}) \odot \rho^3(\textbf{b})  \odot \rho^2(\textbf{c})  \odot \rho^1(\textbf{z})  \odot \rho^0(\textbf{e}). 
\end{equation*}

Another feature of both sequence mappings is that the permutation operation distributes over both binding and superposition operations.
This means that in both mappings the whole sequence can be shifted relative to the initial position by applying the permutation operation required number of times. 
For example, when applying the permutation operation $3$ times to $\textbf{s}$ for $(a,b,c,d,e)$ we obtain: 
\noindent
\begin{equation*}
\rho^{3}(\textbf{s}) = \rho^7(\textbf{a}) + \rho^6(\textbf{b}) + \rho^5(\textbf{c}) + \rho^4(\textbf{d}) + \rho^3(\textbf{e}). 
\end{equation*}
\noindent
Thus, $\rho^{3}(\textbf{s})$ is the shifted version of the original sequence. 
This feature  can be used for sequence concatenation. 
For example, to concatenate   $(a,b,c,d,e)$ and $(x,y,z)$, we can use already calculated $\textbf{s}$ for $(a,b,c,d,e)$ as follows: 
\noindent
\begin{equation*}
\begin{split}
&\rho^{3}(\textbf{s}) + \rho^2( \textbf{x}) + \rho^1(\textbf{y}) + \rho^0(\textbf{z})  =  
\rho^7(\textbf{a}) + \rho^6(\textbf{b}) + \\ & + \rho^5(\textbf{c}) + \rho^4(\textbf{d}) + \rho^3(\textbf{e}) + \rho^2( \textbf{x}) + \rho^1(\textbf{y}) + \rho^0(\textbf{z}).
\end{split}
\end{equation*}
\noindent
This feature was applied in~\cite{BidTrans14} for searching the best alignment (shift) of two sequences that results in the maximum number of coinciding elements.
Other examples of using distributed representation of sequences include modeling human perception of word similarity~\cite{HannaganHolographic2011,CohenOrthogonality2013,RachkovskijEquivariant2021, RachkovskijRecursive2022}, modeling human working memory~\cite{ChooSerialOrderRecall2010, BlouwWordOrder2013, KellyDeclarativeMemory2020, GosmannUnifiedSpikingModel2020, ReimannModellingSerial2021,CalmusBindingNeurobiologically2019}, DNA string matching~\cite{KimEfficientDNA2020}, and
spell checking~\cite{KleykoPermuted2016,RachkovskijEquivariant2021}.

An evident limitation of the above mappings is that due to the usage of a random permutation $\rho$, elements of the sequence in the nearby positions are dissimilar (even if the elements are the same). 
A possible way to handle this limitation is by using locality-preserving representations to encode positions; see some proposals in~\cite{CohenOrthogonality2013, RachkovskijEquivariant2021, schlegel2022hdc, RachkovskijRecursive2022}.
Generally, for a given problem, it might be useful to consider alternative representations that bind element and position hypervectors.
Another limitation is that the representations of the element's order here used hypervector transformation by the permutation corresponding to its absolute position in a sequence.
Thus, the resultant hypervector does not reflect the information about, e.g., successor/predecessor information. 
Some ways of using relative positions when representing sequences in HDC/VSA are investigated in~\cite{HannaganHolographic2011}.

\subsubsection{\textit{n}-gram statistics}
\label{sect:ngrams}

The $n$-gram  statistics of a sequence $S$ is the histogram of all length $n$ substrings occuring in the sequence.  
The mapping of $n$-gram statistics to a single hypervector was presented in, e.g.,~\cite{RIJHK2015}, and includes two steps using 
the primitives above: First, forming hypervectors of $n$-grams, and second, forming a hypervector of the frequency distribution.
The hypervectors of $n$-grams are formed as in Section~\ref{sect:seq} using the chain of binding operations, i.e., each $n$-gram is treated as an $n$-tuple. 
The hypervectors of $n$-grams and their counters are then used to form a single hypervector for the frequency distribution as in Section~\ref{sect:freq}.
Thus, in essence this is a frequency distribution with compound symbols.

The advantage of this mapping is that in order to create a representation for any $n$-gram, we only need to use a single item memory and several simple operations where the number of operations is proportional to $n$. 
In other words, with the fixed amount of resources the appropriate use of operations allows forming a combinatorially large number of new representations.

The mapping, obviously, inherits the limitations of its intermediate steps.
That is, due to the usage of the chain of binding operations (Section~\ref{sect:seq}) similar $n$-grams are going to be mapped to dissimilar hypervectors (assuming that all $n$-gram are assigned with random seed hypervectors). 
And due to the representation of the frequency distribution (Section~\ref{sect:freq}), the retrieved values of individual $n$-grams can be either overcount or undercount.

This mapping has been found useful in several applications: in language identification~\cite{RIJHK2015},  news article classification~\cite{Rasti2016}, and biosignal processing~\cite{HDGestureIEEE} that leveraged its  hardware-friendliness~\cite{RahimiLPHD}.
Distributed representations were also used to untie the dimensionality of the hypervector representing $n$-grams statistics from the possible number of $n$-grams, which grows exponentially with $n$ and would dictate the size of a localist representation of the $n$-grams statistics.
The same property was also leveraged for constructing more compact neural networks using the distributed representation of $n$-grams statistics as their input~\cite{PSI19,HyperEmbed,BandaragodaTrajectoryTraffic2019}.

\subsubsection{Graphs}
\label{sect:graphs}

\begin{figure}[tb]
\centering
\includegraphics[width=0.8\columnwidth]{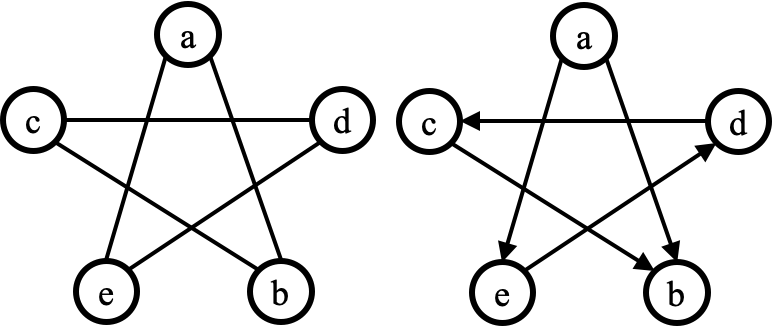}
\caption{
An example of an undirected and directed graphs with $5$ nodes. In the case of the undirected graph, each node has two edges. 
}
\label{fig:graph}
\end{figure}

A graph (denoted as $G$) consists of vertices and edges.
Edges can either be undirected or directed. 
Fig.~\ref{fig:graph} presents examples of both directed and undirected graphs. Following earlier work on graph representations with hypervectors, e.g., in~\cite{CDT2001, GrIsom, GuoGraph2016},  we consider the following very simple mapping of graphs into hypervectors~\cite{GrIsom}.
A random hypervector is assigned to each vertex of the graph, according to Fig.~\ref{fig:graph} vertex hypervectors are denoted by letters (i.e., $\textbf{a}$ for vertex ``a'' and so on).
An edge is represented via the binding operation applied to the hypervectors of the connected vertices, for instance, the edge between vertices ``a'' and ``b'' is represented as $\textbf{a} \odot \textbf{b}$. 
The whole graph $G$ is represented simply as the superposition of hypervectors representing all edges in the graph, e.g., the undirected graph in Fig.~\ref{fig:graph} is: 
\noindent
\begin{equation*}
\textbf{g} = \textbf{a} \odot \textbf{b} + \textbf{a} \odot \textbf{e} + \textbf{b} \odot \textbf{c} + \textbf{c} \odot \textbf{d} + \textbf{d} \odot \textbf{e}.  
\end{equation*} 
\noindent 
Note that if an edge is represented as the result of binding of two hypervectors for vertices, it has no information about the direction of the edge and, therefore, the representation above will not work for directed graphs. 
The direction of an edge can be added applying a permutation to the hypervector of the incidental node, the directed edge from the vertex ``a'' to ``b'' in Fig.~\ref{fig:graph} is represented as $ \textbf{a} \odot \rho(\textbf{b})$.
Note that this is just the mapping of an ordered pair ($2$-tuple in this case) based on binding described in Section~\ref{sect:seq}.
Thus, the directed graph in Fig.~\ref{fig:graph} is represented by the hypervector:
\noindent
\begin{equation*}
\begin{split}
\textbf{g} = &  
\textbf{a} \odot \rho(\textbf{b}) + 
\textbf{a} \odot \rho(\textbf{e}) + 
\textbf{c} \odot \rho(\textbf{b}) + \\
&+\textbf{d} \odot \rho(\textbf{c}) + 
\textbf{e} \odot \rho(\textbf{d}).
\end{split}
\end{equation*}
\noindent
The described graph representations $\textbf{g}$ can be queried for the presence of a particular edge. For graphs that have the same vertex hypervectors, the inner product is a measure of the number of overlapping edges.
When it comes to the usage of the described mappings, 
\cite{GrIsom} propose an HDC/VSA based algorithm for graph matching. For two graphs for which the correspondence between their vertices is unknown, graph matching finds the best match between the vertices so that the graph similarity can be assessed.
In~\cite{MaHolisticMemoriz2018}, a similar mapping is applied on the task of inferring missing links of knowledge graphs.   
The mapping can also be extended to the case when some of its part is learned from the training data; in \cite{NickelHoloGraph2016} representations of knowledge graphs are constructed with hypervectors of nodes and relations that are learned from data.

The described mappings have a number of limitations.
First, they do not work for sparse graphs in which vertices can be entirely isolated because those vertices are not represented at all. 
One way to address it is by also superimposing to $\textbf{g}$ the hypervectors representing the vertices, or to keep a separate hypervector with the superposition of all the vertices.
Another limitation is that one could come up with operations that cannot be done directly on the representation in $\textbf{g}$.
One example of such an operation is the computation of composite edges in a directed graph (see details in~\cite{QiuGraph2022}).

\subsubsection{Binary trees}
\label{sec:trees}

\begin{figure}[tb]
\centering
\includegraphics[width=0.7\columnwidth]{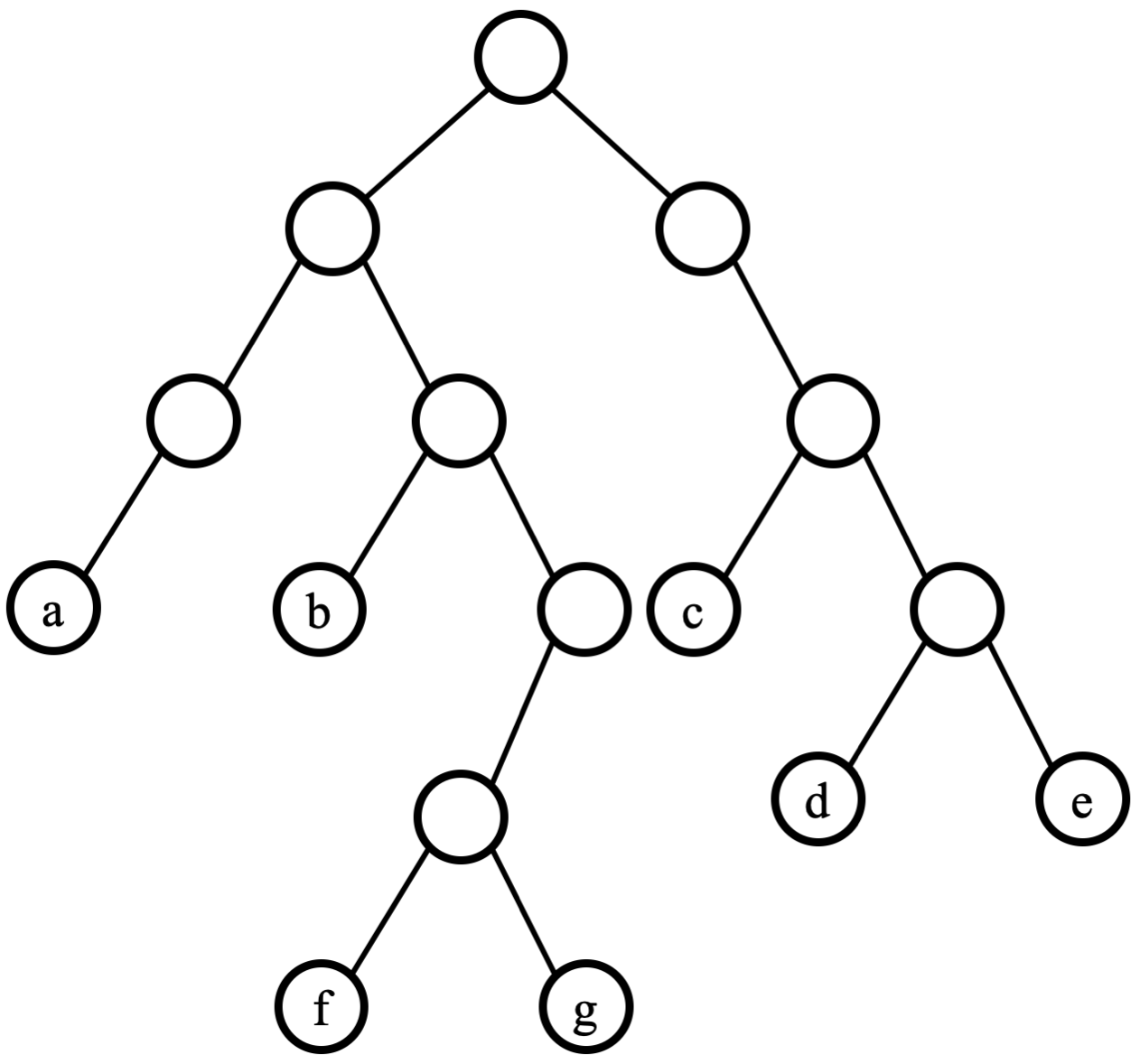}
\caption{
An example of a binary tree from~\cite{ResPart1} where the leaves are different symbols from the alphabet. 
}
\label{fig:BT}
\end{figure}

A binary tree is a well-known data structure where each node has at most two children: the left child and the right child. 
Fig.~\ref{fig:BT} depicts an example of a binary tree, which will be used to demonstrate the mapping of such a data structure into a single hypervector.
We describe a mapping process~\cite{ResPart1} that involves all the three basic HDC/VSA operations and two item memories. 
One item memory stores two random hypervectors corresponding to roles for the left child (denoted as $\textbf{l}$) and the right child (denoted as $\textbf{r}$).
Another item memory stores random hypervectors corresponding to symbols of the alphabet, which are associated with the leaves. 
The example below uses letters so these hypervectors are denoted correspondingly (i.e., $\textbf{a}$ for ``a'' and so on).

The permutation operation is used to create a unique hypervector corresponding to the association of the left or right child with its level in the tree. 
For example, the left child at the second level is represented as $\rho^2(\textbf{l})$.
In general, the level of the node equals the number of times the permutation operation is applied to its role hypervector. 

The chain of the binding operations is used to create a hypervector corresponding to the trace from the tree root to a certain leaf, associated with the leaf's symbol.
For instance, to reach the leaf ``a'', it is necessary to traverse three left children.
In terms of HDC/VSA, this trace will be represented as: $\textbf{a} \odot  \textbf{l} \odot  \rho(\textbf{l}) \odot  \rho^2(\textbf{l})$. 
In such a way, traces to all leaves can be represented.

Finally, the superposition operation is used to combine hypervectors of individual traces in order co create a single hypervector (denoted as $\textbf{t}$) corresponding to the whole binary tree. 
Combining all steps together, the single hypervector for the tree depicted in Fig.~\ref{fig:BT} will then look like:
\noindent
\begin{equation*}
\begin{split}
\textbf{t} = &  
\textbf{a} \odot  \textbf{l} \odot  \rho(\textbf{l}) \odot  \rho^2(\textbf{l}) + \\
&
+\textbf{b} \odot  \textbf{l} \odot  \rho(\textbf{r}) \odot  \rho^2(\textbf{l}) + \\ 
&
+\textbf{c} \odot  \textbf{r} \odot  \rho(\textbf{r}) \odot  \rho^2(\textbf{l}) + \\
& 
+\textbf{d} \odot  \textbf{r} \odot  \rho(\textbf{r}) \odot  \rho^2(\textbf{r}) \odot  \rho^3(\textbf{l}) + \\
&
+\textbf{e} \odot  \textbf{r} \odot  \rho(\textbf{r}) \odot  \rho^2(\textbf{r}) \odot  \rho^3(\textbf{r}) + \\
&
+\textbf{f} \odot  \textbf{l} \odot  \rho(\textbf{r}) \odot  \rho^2(\textbf{r}) \odot  \rho^3(\textbf{l}) \odot  \rho^4(\textbf{l}) + \\
&
+\textbf{g} \odot  \textbf{l} \odot  \rho(\textbf{r}) \odot  \rho^2(\textbf{r}) \odot  \rho^3(\textbf{l}) \odot  \rho^4(\textbf{r}).
\end{split}
\label{eq:tree:repr}
\end{equation*}
\noindent
Thus, the information about the tree in Fig.~\ref{fig:BT} is stored in a distributed way in the compound hypervector $\textbf{t}$, which in turn can be queried with HDC/VSA operations. 
For example, given a trace of children, we can extract the symbol associated with the leaf at this trace. 
Assume that the trace is right-right-left, then its hypervector is $\textbf{r} \odot  \rho(\textbf{r}) \odot  \rho^2(\textbf{l})$.
This hypervector can be unbound from $\textbf{t}$ as: 
\noindent
\begin{equation*}
\textbf{t} \odot (\textbf{r} \odot  \rho(\textbf{r}) \odot  \rho^2(\textbf{l})) = \textbf{c} + \mathrm{noise}.
\label{eq:tree:unbind}
\end{equation*}
\noindent
The result is $ \textbf{c} + \mathrm{noise}$ because $\textbf{r} \odot  \rho(\textbf{r}) \odot  \rho^2(\textbf{l})$ cancels out itself in \textbf{t} and, thus, releases $ \textbf{c}$, which was bound with this trace. 
Since there were other terms in the superposition $\textbf{t}$, they act as crosstalk noise for $\textbf{c}$, hence, denoted as $\mathrm{noise}$.
Thus, when $ \textbf{c} + \mathrm{noise}$  is presented to the item memory, the item memory is expected to return $\textbf{c}$ as the closest alternative, with high probability. 
The inverse task of querying  the trace with a given leaf symbol is more challenging because the resultant hypervector corresponds to a chain of binding operations, e.g., for $\textbf{c}$ we get:    
\noindent
\begin{equation*}
\textbf{t} \odot  \textbf{c} = \textbf{r} \odot  \rho(\textbf{r}) \odot  \rho^2(\textbf{l})  + \mathrm{noise}.
\label{eq:tree:unbind:symb}
\end{equation*}
\noindent
In order to interpret the resultant hypervector one has to query all hypervectors corresponding to all possible traces in a binary tree of the given depth,
where the number of the traces grows exponentially with the depth of the tree. 
This is a significant limitation of the representation.
This limitation can, however, be addressed in part by the resonator network~\cite{ResPart1, ResPart2}  (see Section~\ref{sect:probing}).

We do not cover the details of factoring trees with the resonator network here, but the interested readers are referred to Section 4.1 in~\cite{ResPart1}. 
It should, of course,  be noted that resonator networks are not limitless in their capabilities, since as reported in~\cite{ResPart2}, for the fixed dimensionality of hypervectors their capacity decreases with the increased number of factors (i.e., tree depth in this case).
Nevertheless, they still seem to be the best alternative to tackle the problem (cf. Fig.~3 in~\cite{ResPart2}) -- their search space scales quadratically with $N$.

The presented mapping is, of course, not the only possible way to represent binary trees. 
For example, in \cite{Kanerva19} it was proposed to use two different random permutations for representing nested structures. 
This mechanism can be applied to trees as well by using these  different random permutations  instead of $\textbf{l}$ and $\textbf{r}$.

Last but not least, note that the mapping for binary trees can be easily generalized to trees with nodes having more than two children by superimposing additional role hypervectors in the item memory. 
Also, filler hypervectors  for the leaves do not have to be seed hypervectors -- they could represent any compound structure.

\subsubsection{Stacks}
\label{sect:stack}

A stack is a memory in which elements are written or removed in a last-in-first-out manner.
At any given moment, only the top-most element of the stack can be accessed and elements written to the stack before are inaccessible until all later elements are removed. There are two possible operations on the stack: writing (pushing) and removing (popping) an element.
The writing operation adds an element to the stack -- it becomes the top-most one, while all previously written elements are ``pushed down''.
The removing operation allows reading the top-most element of the stack.
Once read, it is removed from the stack and the remaining elements are moved up.

HDC/VSA-based representations of a stack were proposed in~\cite{StewartSentenceProcessing2014} and~\cite{UCBHD_FSA}. 
The representation of a stack is essentially the representation of a sequence with the addition of an operation that always moves the top-most element to the beginning of the sequence.
For example, if ``\textit{d}'', ``\textit{c}'', and ``\textit{b}'' were successively added to the stack than the hypervector for the current state of the stack is:
\noindent
\begin{equation*}
\textbf{s} = \textbf{b} + \rho(\textbf{c}) + \rho^2(\textbf{d}).
\end{equation*}
\noindent
Thus, the pushing operation is implemented as the concatenation of two sequences (i.e., a new element to be written and the current state of the stack) using their corresponding hypervectors (Section~\ref{sect:seq}).
In particular, the hypervector of the newly written element is added to the permuted hypervector of the current state of the stack. 
For instance, writing ``\textit{a}'' to the current  state $\textbf{s}$ is done as follows:
\noindent
\begin{equation*}
\textbf{s} = \textbf{a} + \rho(\textbf{s}) = \textbf{a} + \rho(\textbf{b}) + \rho^2(\textbf{c}) + \rho^3(\textbf{d}).
\end{equation*}
\noindent
The popping operation includes two steps. 
First, $\textbf{s}$ is probed with the item memory of elements' hypervectors in order to get the closest match for the seed hypervector of the top-most element. 
Once the hypervector of the top-most element is identified (e.g., $\textbf{a}$ in the current example), it is removed from the stack and the hypervector representation of the stack with the remaining elements is moved back by the permutation operation:
\noindent
\begin{equation*}
\begin{split}
\rho^{-1}(\textbf{s}-\textbf{a}) & =  \rho^{-1}(\rho(\textbf{b}) + \rho^2(\textbf{c}) + \rho^3(\textbf{d}))= \\ 
&= \textbf{b} + \rho(\textbf{c}) + \rho^2(\textbf{d}).
\end{split}
\end{equation*}

When it comes to limitations of this representation, there are several things to keep in mind. 
First, the popping operation will not work well if the hypervector representing the stack is normalized after each writing operation, so the operations described above assume that $\textbf{s}$ is not normalized.
Second, the size of the stack that can be retrieved reliably from $\mathbf{s}$ depends on the dimensionality of $\mathbf{s}$. 
Third, if the alphabet of symbols that can be stored in the stack is large, then the probing process for the popping operation might be a computationally demanding step. 
Fourth, if the stack is going to store compound hypervectors, then the popping operation would be more complicated as it either would require the item memory storing all compound hypervectors (this option quickly expand the item memory) or would need to incorporate retrieval procedure assuming the knowledge of the structure of the compound hypervectors so that they could be parsed. 

The main foreseen application of the presented representation is within some control structures as a part of HDC/VSA systems. 
For example, it was used in~\cite{UCBHD_FSA} in a proposal for implementing stack machines and in~\cite{StewartSentenceProcessing2014} as a part of HDC/VSA implementation of a general-purpose left-corner parsing with simple grammars.

\subsubsection{Finite-state automata}
\label{sect:fsa}

A deterministic finite-state automaton is an abstract computational model; it is specified  
by defining a finite set of states, a finite set of allowed input symbols, a transition function, the start state, and a finite set of accepting states. 
The automaton is always in one of its possible states.
The current state can change in response to an input. The current state and input symbol together uniquely determine the next state of the automaton. Changing from one state to another is called a transition.
The transition function defines all transitions in the automaton.

\begin{figure}[t]
\centering
\includegraphics[width=0.8\columnwidth]{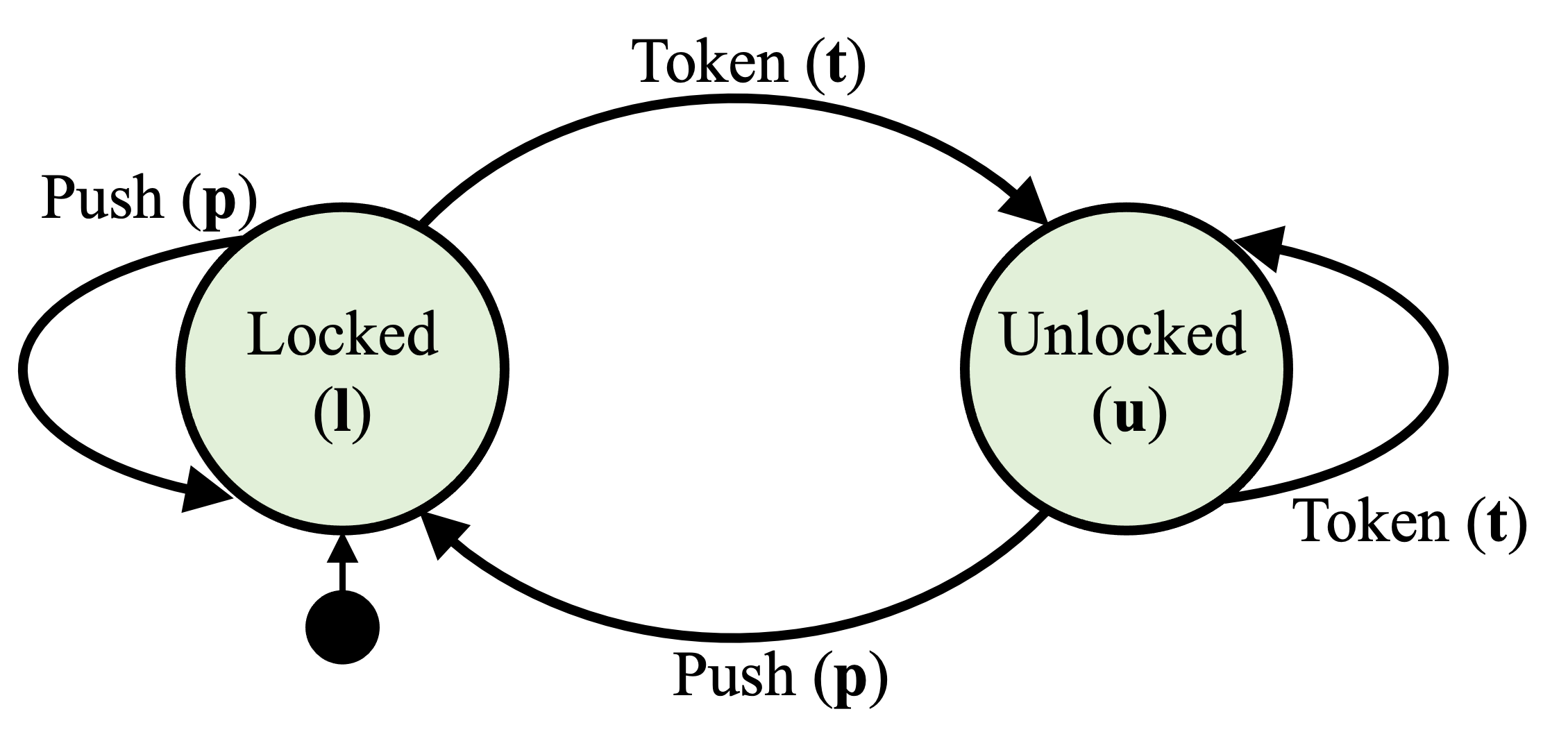}
\caption{
An example of a state diagram of a finite-state automaton modelling the control logic of a turnstile. 
It has two states. The start state is depicted by the arrow pointing from the black circle.
}
\label{fig:fsa}
\end{figure}

Fig.~\ref{fig:fsa} presents an intuitive example of a finite-state automaton, the control logic of a turnstile. 
The set of states is \{ ``Locked'', ``Unlocked'' \} and the set of input symbols is \{ ``Push'', ``Token'' \}. 
The transition function can be easily derived from the state diagram in Fig.~\ref{fig:fsa}. 

HDC/VSA-based implementations of finite-state automata were proposed in~\cite{HD_FSA, UCBHD_FSA}.
Similar to binary trees, the mapping involves all three HDC/VSA operations and requires two item memories. 
One item memory stores seed hypervectors corresponding to the set of states (denoted as $\textbf{l}$ for ``Locked'' and $\textbf{u}$ for ``Unlocked''). 
Another item memory stores seed hypervectors corresponding to the set of input symbols (denoted as $\textbf{p}$ for ``Push'' and $\textbf{t}$ for ``Token''). 
The hypervectors from the item memories are used to form a single hypervector (denoted as $\textbf{a}$), which represents the transition function. 
Note that the state diagram of a finite-state automaton is essentially a directed graph in which each edge has an input symbol associated with it. 
Therefore, the mapping for the transition function is very similar to the mapping of the directed graph in Section~\ref{sect:graphs}. 
The only difference is that the binding of the hypervectors for the vertices, (i.e., states) involves, as an additional factor, the hypervector for the input symbol, which causes the transition.
For example, the transition from  ``Locked'' state to ``Unlocked'' state, contingent on receiving ``Token'', is represented as: 
\noindent
\begin{equation*}
\textbf{t} \odot \textbf{l} \odot \rho(\textbf{u}).
\end{equation*}
\noindent
Given the distributed representations of all transitions of the automaton, the transition function $\textbf{a}$ of the automaton  is represented by the superposition of the individual transitions: 
\noindent
\begin{equation*}
\textbf{a} =   
\textbf{p} \odot \textbf{l} \odot \rho(\textbf{l}) + 
\textbf{t} \odot \textbf{l} \odot \rho(\textbf{u}) + 
\textbf{p} \odot \textbf{u} \odot \rho(\textbf{l}) + 
\textbf{t} \odot \textbf{u} \odot \rho(\textbf{g}).
\end{equation*}
\noindent

In order to calculate the next state, we query $\textbf{a}$ with the binding of the hypervectors of the current state and input symbol followed by the inverse permutation operation applied to the result. 
Calculated in this way, the result is the noisy version of the hypervector representing the next state. 
For example, if the current state is $\textbf{l}$ and the input symbol is $\textbf{p}$ then we have:
\noindent
\begin{equation*}
\rho^{-1}(\textbf{a} \odot \textbf{p}  \odot \textbf{l} )  =  \textbf{l} +  \mathrm{noise}. 
\end{equation*}
\noindent
As usual, this hypervector should be passed to the item memory in order to retrieve the noiseless seed hypervector $\textbf{l}$.

The same mapping can be used to create a hypervector representing a nondeterministic finite-state automaton~\cite{RabinFinite1959}. The main difference from deterministic finite-state automata is that the nondeterministic finite-state automaton can reside simultaneously in several of its states. The transitions do not have to be uniquely determined by their current state and input symbol, i.e., there can be several valid transitions from a given current state and input symbol.
The nondeterministic finite-state automaton can assume a so-called generalized state, defined as a set of the automaton's states that are simultaneously active. The generalized state corresponds to a hypervector representing the set of the currently active states with (\ref{eq:set}).
Similar to the deterministic finite-state automata, the hypervector for the generalized state is used to query the automaton to get a hypervector for next generalized state.
This corresponds to a parallel execution of the automaton from all currently active states. 
It should also be noted that in the case of the nondeterministic finite-state automaton, due to the potential presence of several active states, the cleanup procedure (Section~\ref{sect:probing}) has to search for several nearest neighbors. 
Please see Section~\ref{sec:string:search} for an example of such a procedure.


\begin{figure}[tb]
\centering
\includegraphics[width=1.0\columnwidth]{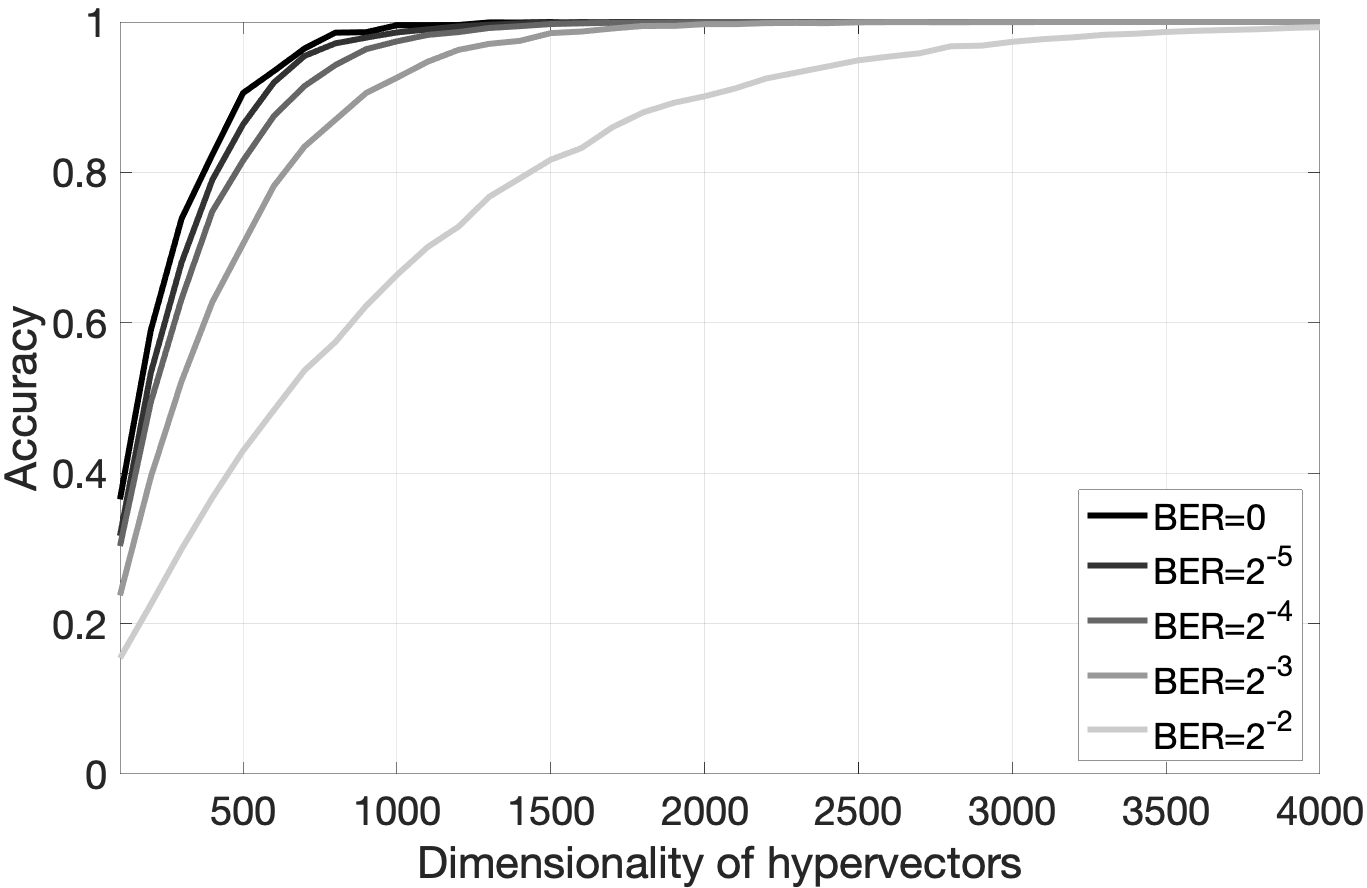}
\caption{
Average accuracy of the recall of the next state of the automaton from $\textbf{a}$, which was bipolarized, against the dimensionality of hypervectors ($N \in [100, 4000]$, with step $100$).
The results were obtained over $50$ random initializations of the item memories. 
For each initialization, $1,000$ transitions (chosen randomly) were performed. 
For each transition function, noise added to $\textbf{a}$ was also generated at random. 
Bit Error Rates were in range $0.0312$--$0.2500$, Bit Error Rate is defined as the percentage of bits (here dimensions) that have errors relative to the total number of bits.
}
\label{fig:fsa:recall}
\end{figure}

In the next subsection, we will see an example of how to compute with hypervectors representing automata, but the most obvious application of the presented representation is to execute the automaton in the presence of noise in hypervectors. 
Fig.~\ref{fig:fsa:recall} presents the accuracy of the correct recall of a next state from a bipolarized hypervector representing an automaton with $22$ states and $29$ symbols. 
The figure shows how the accuracy changed with the dimensionality of hypervectors for different values of noise in $\textbf{a}$.
As expected, we see that for every amount of noise, there is eventually a dimensionality that allows a perfect recall -- an elegant property that can be simply leveraged for executing a deterministic behavior in a very stochastic environment.

While currently there are not many HDC/VSA applications that use finite-state automata (but we will see one in Section~\ref{sec:string:search}), there is a potential in such a mapping as it naturally allows using HDC/VSA as a medium for executing programs that can be formalized via automata.
Moreover, the primitives for stacks and finite-state automata can be combined to create richer computational models such as deterministic pushdown automata or stack machines; see, e.g.,~\cite{UCBHD_FSA} for a sketch of a stack machine operating with hypervectors.  
An alternative representation for pushdown automata and context-free grammars has been recently presented in~\cite{VSAGrammars}.

Finally, it should be noted that the presented mapping is designed for executing an automaton, however, it is limited in the sense that it cannot be used directly to modify it or to perform composition operations (e.g., combining it with another automaton).

\subsubsection{Deeper hierarchies}

Finally, it is important to touch upon constructing data structures encoding deep hierarchies. 
In the previous subsections, we concentrated mainly on data structures with a single level hierarchy. 
In fact, this is what most of the current studies in the area used.
Therefore, we will not go into technical details of existing proposals.
HDC/VSA, however, are well-suited for representing many levels of hierarchy and representation of hierarchical data structures was a part of the original motivation right from the start (see, e.g.,~\cite{PlateThesis}).
The representation of binary trees in Section~\ref{sec:trees} can already be seen as a hierarchy, since a tree has several levels and the representation should be able to discriminate between different levels. 
In the presented mapping, this was done using powers of permutation to protect different levels of hierarchy. 
This can be done in some other ways by, e.g., assigning special role hypervectors for each level. 
Currently, the usage of representations for hierarchies in HDC/VSA is relatively uncommon. 
We mainly attribute this fact to the nature of applications which are being explored, rather than to the capabilities of HDC/VSA. 
The use-cases, which relied on the representation the hierarchical representations, are representation of analogical episodes~\cite{PlateBook, RachkovskijAnalogical2004}, 
distributed orchestration of workflows~\cite{SimpkinHDWorkflow2019}, and representation of hierarchies in WordNet concepts~\cite{CrawfordKnowledge2016}.
It has also been argued that the representation of hierarchical data structures via HDC/VSA is an important feature for modular learning where  modules at different levels of hierarchy can communicate with such representations~\cite{RachkovskijWorldModel2013, GhaziRecursive2019}.
Finally, there is a recent proposal that suggests that the JSON format with several levels of hierarchy can be represented in hypervectors~\cite{GallantVSAJSON2022}.


\subsection{Computing in superposition with HDC/VSA}
\label{sect:comp:super}

\subsubsection{Simple examples of computing in superposition}

A well-known data structure -- Bloom filter~\cite{Bloom1970space} -- is the simplest case of computing in superposition. 
Bloom filter is a sketch as a fixed-size memory footprint is used to represent a set of elements.
A Bloom filter encodes a set as a superposition of its elements' sparse binary vectors, which, in essence, corresponds in HDC/VSA to a compound hypervector representing sets.
Thus, Bloom filter directly corresponds to the primitive for representing sets as described in Section~\ref{sect:sets}.
With Bloom filters, the algorithm for searching an element in a set is a single operation of comparing the similarity of the distributed representation of the query element to the Bloom filter instance. 
In other words, all elements of the set are tested in one shot, i.e., the search is performed as a computation in superposition.
It enables solving the approximate membership query task instantaneously. This illustrates a simple instance of computing in superposition.
Bloom filters are highly specialized for one particular task. 
In contrast, HDC/VSA constitute a broad framework for computing in superposition, containing Bloom filters as a subclass~\cite{ABF}.
We have already seen other examples in Section~\ref{sect:primitives} for computing in superposition with HDC/VSA, such as the primitives for recursive construction of sequence representations (see equations (\ref{eq:perm_seq}) and (\ref{eq:perm_seq1})) and in Section~\ref{sect:tup} the forming of a representation for the cross product of two sets via a single binding operation. 
In these examples, the distributivity of HDC/VSA operations (see Section~\ref{sec:VSA:properties}) played an important role.

\begin{figure}[t]
\centering
\includegraphics[width=1.0\columnwidth]{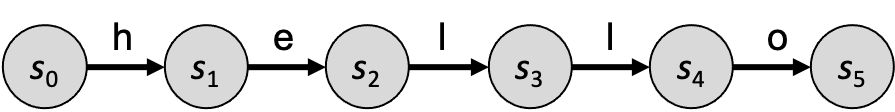}
\caption{
The automaton for the base string ``hello''. 
}
\label{fig:string:automata}
\end{figure}

\subsubsection{Computing in superposition for substring search}
\label{sec:string:search}

Finding a substring within a larger string is a standard computer science problem with numerous algorithms (e.g., \cite{Boyer_string,Karp_string,Knuth_string}) that have a linear complexity on the total length of the base and the query strings.
Recently, an algorithm based on nondeterministic finite-state automata was formulated with HDC/VSA~\cite{PashchenkoSubstring2020}. 
It nicely demonstrates how HDC/VSA can solve computer science problems, so we briefly explain it here.

Each position of a symbol in the base string is modeled as a unique state of the nondeterministic finite-state automaton $S=\{s_0, s_1, s_2, \ldots, s_n\}$. 
For example, the string ``hello'' generates an automaton with six states: $s_0$ through $s_5$. The transitions between states are defined by the base string's (denoted $B$) symbols $b_i$ from $B=\{b_1, b_2, \ldots, b_n\}$. 
Fig.~\ref{fig:string:automata} illustrates the automaton for the string ``hello.''
The nondeterministic finite-state automaton is then defined by tuple $<S, s_0, B, T>$, where $s_0$ is the start state of the automaton  and $T$ is the set of transition tuples of the form $t_i=<s_{i-1}, b_i, s_i>$, where $s_{i-1}$ and $s_i$ are the start and end states of a particular transition caused by symbol $b_i$. 
The elements of sets $B$ and $S$ are represented by i.i.d. random hypervectors (denoted in bold). 
The hypervector $\boldsymbol\beta$ of the automaton for the base string is constructed as (cf. Section~\ref{sect:fsa}):
\noindent
\begin{equation}
\boldsymbol\beta =\sum_{i=1}^{|B|} \textbf{s}_{i-1} \odot \textbf{b}_i \odot \rho^1(\textbf{s}_i).   
\end{equation}
\noindent
Thus, $\boldsymbol\beta$ is the superposition of all the automaton's transitions caused by sequential input of symbols of the base string. 
Note that this representation corresponds to the primitive for the finite-state automata as described in Section~\ref{sect:fsa}.

\begin{figure}[t]
\centering
\includegraphics[width=1.0\columnwidth]{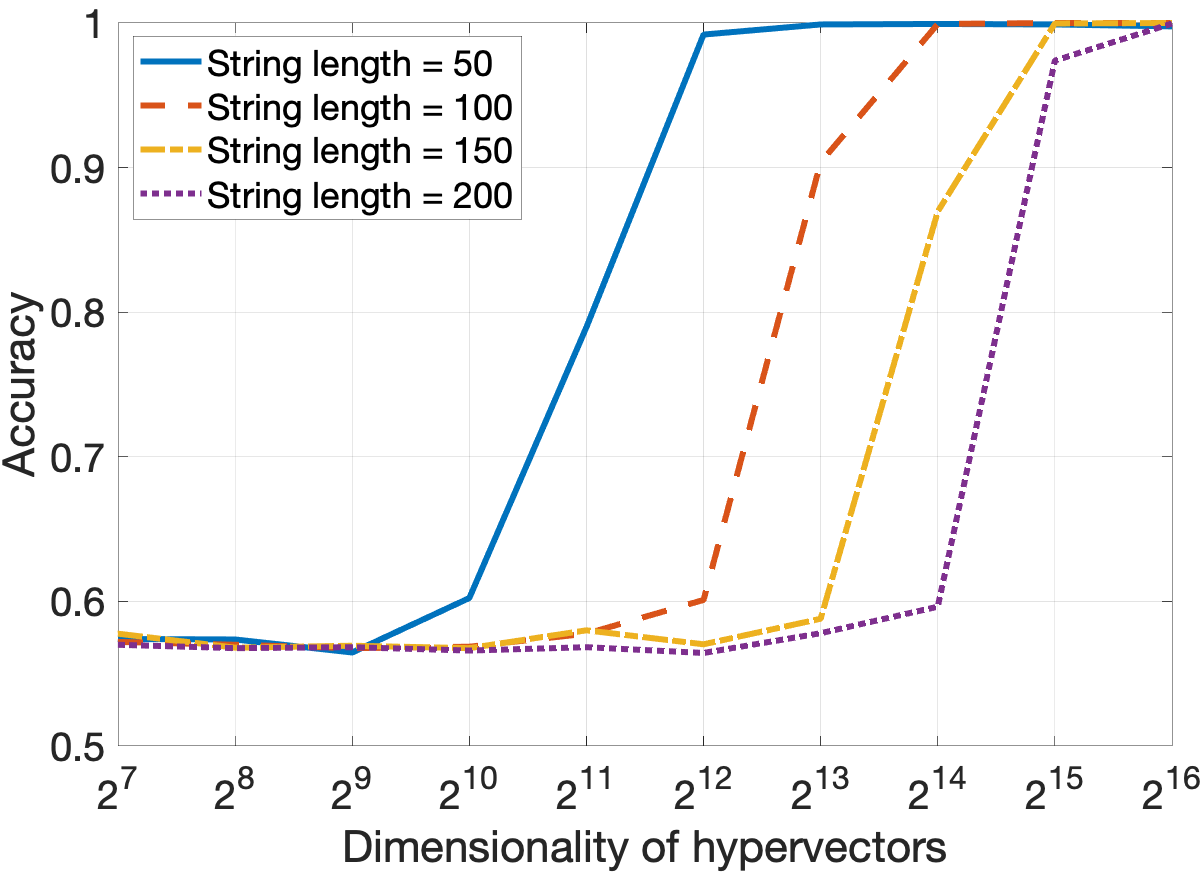}
\caption{
Search of a substring in superposition with HDC/VSA using the modified algorithm from~\cite{PashchenkoSubstring2020}.
The length of a substring was fixed to $30$.
The reported values were averaged over $30$ simulations.
}
\label{fig:SS:search}
\end{figure}

The algorithm for finding whether a query string $Q=\{q_1, \ldots, q_l \}$ is a part of the base string $B$ is a sequential retrieval of the next state of automaton $\boldsymbol\beta$, for each symbol of the query string $q_j$.
In terms of hypervectors, this is: 
\noindent
\begin{equation}
{\textbf{p}}_j=\rho^{-1}( {\textbf{p}}_{j-1} \odot \boldsymbol\beta \odot \textbf{q}_j),
\label{eq:ndfsa}
\end{equation}
\noindent
where ${\textbf{p}}_j$ denotes the hypervector that includes the hypervector(s) of the next generalized automaton state (given symbol $q_j$), as well as crosstalk noise. 
Equation (\ref{eq:ndfsa}) is also a primitive from Section~\ref{sect:fsa}.
Note, however, that generalized state may include one or several states $s_i$. 
The set of valid (i.e., permitted) previous generalized states is initialized as $\textbf{p}_0=\sum_{ s_i \in S}\textbf{s}_i$, which is a superposition of all the states of the base string. 
Since the operation in (\ref{eq:ndfsa}) is performed on the superimposed set of all states, it is qualified as computing in superposition.

While the algorithm presented in~\cite{PashchenkoSubstring2020} works in principle (confirmed experimentally but not reported here), the required dimensionality of hypervectors grows extremely fast with the length of strings since every step of (\ref{eq:ndfsa}) introduces additional crosstalk noise to ${\textbf{p}}_j$.
Crosstalk noise can be reduced by a cleanup procedure on ${\textbf{p}}_j$ after every execution of (\ref{eq:ndfsa}):
\noindent
\begin{equation}
{\textbf{p}}_j=\mathbf{S}\mathbf{S}^{\top}{\textbf{p}}_j,
\label{eq:ndfsa:cleanup}
\end{equation}
\noindent
where $\mathbf{S} \in [N,n+1]$ denotes the item memory storing hypervectors for the unique states of the base string, $S=\{s_0, s_1, s_2, \ldots, s_n\}$.
This primitive uses the idea of projecting predictions back onto the item memory and it was introduced in Section~\ref{sect:probing} as a part of resonator network (see equation~(\ref{eq:resnet:text})).

We simulated the modified algorithm for searching a fixed length query substring ($30$ symbols) in the base string of four different lengths (see Fig.~\ref{fig:SS:search}).  
Average accuracy in $30$ simulation runs is plotted against the varying dimensionality of hypervectors. In in every simulation run, $100$ different random base strings were used. 
In approximately half of the searches, the query substring was present in the base string, so a single simulation run determines the accuracy of correctly detecting when a substring is present and when it is not (thus, the accuracy of a random guess is $0.5$). 
With increasing dimensionality of hypervectors, the accuracy of detecting a substring increases and eventually approaches $1$.
For longer base strings, it would require larger dimensions of the hypervectors to achieve high accuracy. 
Nevertheless, it scales much better than the original algorithm for which we were not able to simulate large enough dimensionalities that would provide reasonable accuracy.

The substring search provides lessons for computing in superposition with HDC/VSA.
Both algorithms use it; the original one requires a large dimensionality to reduce crosstalk sufficiently, while the modified one includes an extra cleanup step to reduce the required dimensionality significantly -- but it also increases the algorithmic complexity. 
In particular, the asymptotic computational complexity of the query algorithm in HDC/VSA operations is $O(|Q|)$ for the original algorithm versus ($O(|Q||B|)$ for the modified algorithm. But in terms of hypervector dimensionality, the original algorithm required much more space than the modified algorithm.
Another consequence of long hypervectors required by the original algorithm is that despite not requiring an extra cleanup step (\ref{eq:ndfsa:cleanup}), the total number of operations would be higher due to much shorter hypervectors used by the modified algorithm. 
Moreover, with appropriate implementation of the HDC/VSA algorithm on parallel hardware, the cleanup step in (\ref{eq:ndfsa:cleanup}) can be parallelized\footnote{
For the sake of fairness, it should be noted that the conventional substring search algorithms could also be parallelized. 
}
using, e.g., in-memory computing architectures with massive phase-change memory devices~\cite{KarunaratneHDAugmented2020}.
When executed on such hardware, the time complexity of the modified algorithm also becomes $O(|Q|)$.\footnote{
Of course, the size of the chip places limitations on the dimensionality of hypervectors and the number of hypervectors in the item memory.
} 
Thus, computing in superposition in HDC/VSA is natural but can require very high dimensionality for managing crosstalk. 
Steps to manage the crosstalk can be added in the algorithm at no compute time costs, if the algorithm is properly mapped on parallel hardware (see, e.g., \cite{KimEfficientDNA2020} for acceleration of DNA string matching with HDC/VSA).

Last, it is important to note that we do not claim that the substring search will be a practically useful application of computing in superposition, since its computational complexity exceeds that of the conventional algorithms optimized for the problem.
However, we think that this example has a didactic value as it clearly demonstrates how the primitives for representing data structures from Section~\ref{sect:primitives} can be connected to a well-known computer science problem. Thus, it serves as an important illustration of the lines along which one should think to utilize computing in superposition.
Below, we elaborate on more practical (but not always explicit) contemporary examples of using computing in superposition.

\subsubsection{Applications of computing in superposition}

In a long-term, we anticipate the resonator networks~\cite{ResPart1,ResPart2} (see Section~\ref{sect:probing}) to become a pivotal mechanism in many solutions based on computing in superposition since they use the idea of removing crosstalk noise from the predictions represented in the superposition.
In particular, we believe that this idea would be important to efficiently solve non-trivial combinatorial search problems.
There are already a couple of proposals for, e.g., scene decomposition~\cite{FradyDisentangling2018} and prime factorization~\cite{KleykoPrimes2022}, but they are yet to be demonstrated at scale.

In a short-term, there is another practical direction for the application of computing in superposition that is already being used to tackle a large problem -- enhancement of capabilities of machine learning algorithms (often neural networks)\footnote{
We additionally review some of these works in the context of connections to hardware realizations of HDC/VSA in Section~\ref{sec:hardware:neural:nets}.
}.
Below, we briefly explain the role computing in superposition plays in approaches proposed within this direction, since, in our opinion, it is a unifying theme that will, hopefully, inspire more approaches for machine learning algorithms enhancement.

A recent connection, introduced in~\cite{FradyFunctions2021, FradyFunctionsNICE2022}, between a method for representation of numeric data as hypervectors~\cite{PlateThesis, WeissOlshausenSpatial16, KomerContinuous2019} and kernel methods allowed representing functions as compound hypervectors of weighted sets (Section~\ref{sect:sets}).
This finding, in turn, allowed one-shot evaluation of kernel machines since the whole model can now be stored in the superposition as a compound hypervector. 
The one-shot evaluation principle was demonstrated on probability density estimation~\cite{FradyFunctions2021, FradyFunctionsNICE2022,FurlongProbability2022}, kernel regression~\cite{FradyFunctions2021, FradyFunctionsNICE2022}, Gaussian processes-based mutual information exploration~\cite{FurlongSSPMI2022}, and rules search in superposition~\cite{herscheNVSA2022}.
The distributed representations of numeric data can also be very useful even without formal links to the kernel methods. 
They can be used to store in superposition multiple locations of interest on a 2D grid that has been shown to be important for, e.g., implementing agent's memory for cognitive maps~\cite{KomerNavigation2020}, navigation in 2D environments~\cite{WeissOlshausenSpatial16,KomerNavigation2020}, and reasoning on 2D images~\cite{WeissOlshausenSpatial16, FradyDisentangling2018, LuFractional2019}.

When it comes to approaches for augmenting neural networks, in~\cite{CheungSuperposition2019}, the weights of multiple deep neural networks trained to solve different tasks were stored jointly in superposition using a single compound hypervector.
This approach addressed the so-called ``catastrophic forgetting'' phenomenon by using a unique random permutation assigned to each task that allow networks to co-exist in the compound hypervector without much interference. 
These permutations were used as keys to extract the corresponding network's weights from the superposition hypervector.
A big leap of such an approach is that new networks can be added gradually into the superposition hypervector without significant degradation of the performance of the previously included networks.

Another approach combining computing in superposition and neural networks was presented in~\cite{WilsonOOD2021}. 
There, activations of network's layers from a single data sample were used in place of value hypervectors. 
They were bound to the corresponding random key hypervectors and all hypervectors of the key-value pairs were aggregated in a single compound hypervector.
Since the compound hypervector simultaneously keeps all the activations, calculating the similarity between two such hypervector corresponds to an aggregate similarity score between two data samples.
This property was leveraged successfully to detect out-of-distribution data.  
In a similar way, in~\cite{NeubertAggregation2021, NeubertPlaceRecognition2021}, activations of multiple neural network-based image descriptors were combined together into a compound hypervector simultaneously representing the aggregated descriptor.
Such hypervectors allowed an efficient image retrieval for visual place recognition task.
A different combination of a neural network and a compound hypervector of the key-value pairs was reported in~\cite{GanesanLearning2021}, where the compound hypervector was used to simultaneously represent the output of a neural network when solving multi-label classification tasks.

From the descriptions above, one can notice a striking pattern -- most of the approaches relied on the primitive for representing sets, in general, and sets of key-value pairs, in particular.  
This is likely because the latter is a simple yet non-trivial data structure. 
We, thus, anticipate that more new approaches can be conceived by expanding to more sophisticated data structures.

\section{Hardware realizations of HDC/VSA}
\label{sect:hardware}

\subsection{HDC/VSA models for different types of hardware}
\label{sect:frameworks}

The computational  primitives of HDC/VSA connect the algorithmic level of Marr's computing hierarchy  (Fig.~\ref{fig:stack}) to the computational level. 
At the same time, a HDC/VSA placed at the algorithmic level also interfaces with the implementation level. While the computational primitives are generic across different HDC/VSA models, the model choice can become critical when it comes to interfacing with a particular physical substrate.\footnote{
It should be noted that there exist subtleties when it comes to computational primitives of different HDC/VSA models (see, e.g.,~\cite{SchlegelVSAComparison2020} for a discussion). 
So, strictly speaking, model choice may not be only influenced by a physical substrate but also by the nature of the task at the computational level.
To put it simply, not all HDC/VSA models are interchangeable.
This is not entirely unexpected since, if a framework can provide tight matches between computation and hardware to enable efficiency, the separation between abstraction and physical realization cannot be perfect.
Thus, for the sake of narration in this section, we focus on the availability of an efficient mapping between some physical substrate and some HDC/VSA model.
}
This suggests a general design pattern when designing a HDC/VSA system to be implemented on emerging hardware: {\it the desired computation is formalized in terms of the generic HDC/VSA computational primitives, and then the specific HDC/VSA variant best suited for this emerging hardware is used in implementing these primitives}.
Here, we describe some of the existing HDC/VSA models and examples of how they can be implemented in different hardware. 
Different HDC/VSA models can be distinguished in terms of the properties of seed hypervectors and corresponding algebraic operations. 

\subsubsection{Dense binary vectors} 
\label{sec:models:dense}
The Binary Spatter Codes~\cite{Kanerva97} model uses dense binary vectors.
Superposition is done by the component-wise majority rule followed by tie-breaking, and binding is by the component-wise XOR. Due to its discrete nature, Binary Spatter Codes is highly suitable for conventional digital application-specific integrated circuit (ASIC). 
The first ASIC design~\cite{RahimiLPHD} was made in 65\,nm CMOS for the language recognition, followed by more programmable designs in 28\,nm~\cite{DattaProcessor2019} and 22\,nm~\cite{AlwaysOn2021}. It has been also mapped on a 28\,nm FD-SOI silicon prototype with four programmable OpenRISC cores operating in near-threshold regime (0.7\,V--0.5\,V)~\cite{HD_PULP}. Overall, in the Binary Spatter Codes model, the hypervectors are stationary and robust, and related binary operations are local and simple. This provides a natural fit for implementing the model on non von Neumann architectures (a.k.a. in-memory computing) using emerging technologies such as carbon nanotube FETs and resistive RAM~\cite{LiHD16,WuNanotube2018,WuCarbonNanotube2018}, and phase-change memory~\cite{MemristorHD19,KarunaratneHDAugmented2020}. Specifically, ~\cite{MemristorHD19} describes how to organize computational memories for storing and manipulating hypervectors whereby the operations are implemented inside, or near, computational memory elements.

\subsubsection{Integer vectors}
The Multiply-Add-Permute model~\cite{MAP}, the HDC/VSA model we have used in the examples so far as the default, employs bipolar (+1s and -1s) hypervectors, component-wise multiplication, and superposition with possible thresholding.
Multiply-Add-Permute model will usually suit the same technologies as Binary Spatter Codes.
For example, it was recently implemented on an FPGA for hand gesture recognition~\cite{MoinWearable2021}.

\subsubsection{Real-valued vectors}
The Holographic Reduced Representation model~\cite{PlateTr} was originally done with $N$-dimensional real-valued hypervectors whose components
are i.i.d. normal with zero mean and $1/N$ variance.  
Superposition is done by the normalized vector sum, and binding is done by circular convolution.
It has been shown how to map real-valued hypervectors onto spiking neurons using the principles from the Neural Engineering Framework~\cite{EliasmithNeural2003} with the help of spike-rate coding.
For example, the Spaun cognitive architecture~\cite{BuildBrain} has been implemented in such a way. 
Most of the studies were done using simulations in Nengo~\cite{BekolayNengo2014}, which is a Python-based package for simulating large-scale spiking neural networks.
Nevertheless, Nengo has compilers to popular neuromorphic platforms such as SpiNNaker and Loihi, therefore, it is straightforward to deploy a model built in Nengo on the neuromorphic platforms.

\subsubsection{Complex vectors} 
In the Fourier Holographic Reduced Representations~\cite{PlateBook}, vector components are random phasors, superposition is by component-wise complex addition followed by normalization, and binding is by component-wise complex multiplication (addition of phasors)~\cite{PlateBook}.
This HDC/VSA model should be suited for implementations on coupled oscillator hardware~\cite{CsabaCoupled2020}, however, we are not aware of any concrete hardware realizations as of yet. 
Another alternative is mapping complex HDC/VSA to the neuromorphic hardware~\cite{Loihi18}, by representing phasors with spike times~\cite{TPAM}. This implementation is particularly interesting because the neuromorphic hardware scales up more easily than the current approaches to coupled oscillator hardware.  However, no neuromorphic implementation of a full complex HDC/VSA has been reported to date.

\subsubsection{Sparse vectors}
\label{sec:models:sparse}

Traditional HDC/VSA models use dense distributed representations. However, sparsity is an important ingredient of energy efficient realizations in hardware. Thus, HDC/VSA models that use sparse representations are important for mapping HDC/VSA operations efficiently onto hardware. We are aware of two such models: Sparse Binary Distributed Representations~\cite{CDT2001,KleykoSDR2016} and Sparse Block-Codes~\cite{Laiho2015, FradySDR2020}.
In the Sparse Binary Distributed Representations model, the hypervectors are sparse patterns without any restrictions on placing the active components, while in Sparse Block-Codes the hypervectors are divided into blocks of the same size (denoted as $K$) with just one single active component per block. 
The Sparse Binary Distributed Representations model was implemented around 1990 in specialized hardware -- ``associative-projective neurocomputers'' ~\cite{RachkovskijTexture1991}. This hardware was designed to operate efficiently with sparse representations~\cite{CDT2001} by using simple bit-wise logical operations and a long word processor with $256$ bits (later with $512$ and $2048$ bits, implemented by Wacom, Japan). 
For cleanup memory, it used Willshaw-like associative memories, following earlier ideas to implement such memory networks~\cite{palm1984parallel} and motivated by theoretical results suggesting high memory capacity \cite{willshaw1969non, palm1980associative, palm1992info, sommer1998bayesian, FrolovWillshaw2002,FrolovTime2006}.
Concerning HDC/VSA with Sparse Block-Codes, in particular with complex-valued sparse vectors, they seem to be the most amenable for implementations on neuromorphic and coupled oscillator hardware.
Currently, there are two proposals for implementing binary Sparse Block-Codes in spiking neural network circuits~\cite{BentSpike2022, RennerBinding2022}.
The proposal in~\cite{RennerBinding2022} has been implemented on Intel's Loihi~\cite{Loihi18} while the one from~\cite{BentSpike2022} has not been realized in hardware yet, but it has been implemented in the Brian 2 simulator~\cite{stimberg2019brian}.

\subsection{Mapping algorithms to hardware}
\label{sec:mapping:to:hardware}

\subsubsection{Hardware implementations of pure HDC/VSA}
How do implementations of HDC/VSA in existing conventional hardware produce gains over conventional machine learning methods?
On a dedicated digital ASIC design, it has been demonstrated that HDC/VSA-based classification can lower the energy by about 2$\times$ compared to a $k$-nearest neighbors classifier for the European language recognition task~\cite{RahimiLPHD}. By running these classifiers on the Nvidia Tegra X2 GPU, HDC/VSA exhibited over 3$\times$ lower energy per prediction~\cite{DattaProcessor2019}. Considering a wide range of biomedical signal classifications, HDC/VSA achieved at least the same level of accuracy compared to the baseline methods running on the conventional programmable hardware, however, at: 2$\times$ lower power compared to the fixed-point SVM for EMG classification on the embedded ARM Cortex M4~\cite{HD_PULP}; 2.9$\times$ lower energy compared to SVM, and over 16$\times$ compared to CNN and LSTM for iEEG classification on the Nvidia Tegra X2~\cite{BurrelloSeizure2019}. More details for this benchmarking is available in~\cite{HDGestureIEEE}. Using PageRank centrality metric, HDC/VSA achieved comparable accuracy with 2$\times$ faster inference compared to the graph kernels and neural networks for graph classifications on the Intel Xeon CPUs~\cite{GraphHD2022}. These improvements are due to the fact that the HDC/VSA-based solutions mostly use basic bit-wise operations, instead of fixed-point or floating point operations.

Another appealing property of HDC/VSA-based solutions is their great robustness, for example, they tolerate 8.8$\times$ higher probability of failures with respect to intermittent hardware errors~\cite{RahimiLPHD}, and 60$\times$ higher probability of failures with respect to permanent hardware errors~\cite{LiHD16}. This robustness makes HDC/VSA ideally suited to the low signal-to-noise ratio and high variability conditions in the emerging hardware as discussed in more detail in~\cite{HDNP17}. Among them, as a large-scale experimental demonstration~\cite{MemristorHD19} of HDC/VSA, it was implemented on $760,000$ phase-change memory devices performing analog in-memory computing with $10,000$-dimensional binary hypervectors for three different classification tasks. The implementation not only achieved accuracies comparable to software implementations---despite the non-idealities in the phase-change memory devices---but also achieved over $6\times$ end-to-end energy saving compared to an optimized digital ASIC implementation~\cite{MemristorHD19}.

The connection of HDC/VSA to spiking neuromorphic hardware is not obvious since all classical HDC/VSA models used abstract connectionist representations, not spikes. However, recent work has demonstrated that representations of a complex HDC/VSA model, Fourier Holographic Reduced Representations~\cite{PlateBook}, can be mapped to spike timing codes~\cite{TPAM}. Although focused just on content-addressable memory, i.e., item memory, this work opens avenues for efficient implementations of full HDC/VSA models on neuromorphic hardware~\cite{DaviesAdvancingLoihi2021}. 
Because neuromorphic hardware often optimizes spike communication for sparse network connectivity, the scaling properties of neuromorphic HDC/VSA will potentially outperform other types of hardware. 
Further, neuromorphic hardware might enable hybrid approaches by integrating HDC/VSA with other computing frameworks.
For instance, an event-based dynamic vision sensor (as a front-end spiking sensor) has been combined with sparse HDC/VSA leading to $10\times$ higher energy efficiency than an optimized 9-layer perceptron with comparable accuracy on an 8-core low-power digital processor~\cite{HerscheDVSCDT2020}.

The results above bring a question worth discussing -- what are the common hardware primitives enabling these gains?
The most common architectural primitives that are observed in the hardware implementations can, actually, be naturally mapped to basic elements (Section~\ref{sect:methods:basic}) and operations (Section~\ref{sect:oper}) of HDC/VSA.
For example, let us consider the implementations of the Binary Spatter Codes model based on phase-change memory devices reported in~\cite{MemristorHD19} and of the Sparse Block-Codes model on spiking neural network circuits described in~\cite{BentSpike2022}. 
The basic hardware primitives lying at the core of these implementations were:
item memory circuit (cf. Fig.~1 in~\cite{MemristorHD19} \& Section~III-A1a in~\cite{BentSpike2022}),
superposition operation circuit (``The complete in-memory HDC system'' in~\cite{MemristorHD19} \& Fig.~2 in~\cite{BentSpike2022}),
binding operation circuit (cf. Fig.~3 in~\cite{MemristorHD19} \& Fig.~4 in~\cite{BentSpike2022}), and
circuit for probing (cf. Fig.~2 in~\cite{MemristorHD19} \& Fig.~3 in~\cite{BentSpike2022}).

The fact that the basic HDC/VSA elements and operations are the most common hardware primitives should not be surprising because, as it was demonstrated in Section~\ref{sect:primitives}, they are the key building blocks of all the computational primitives in the ``HDC/VSA cookbook''.
This implies that given the hardware implementation of the most basic elements, it is possible to construct architectures for compositional primitives that might, e.g., combine usage of several HDC/VSA operations.
This, of course, does not mean that there is no other way to approach hardware implementation of HDC/VSA. 
In fact, there are incentives to design implementations targeting concrete compositional primitives and they were even present in the two above works, e.g., a circuit for representing $n$-grams -- see Fig.~3 in~\cite{MemristorHD19} and a circuit for representing a set of key-value pairs -- see Fig.~5 in~\cite{BentSpike2022}.  
The main incentive for doing so is to increase the efficiency of the implementation since it allows applying, e.g., computational reuse.
A vivid example of such an approach is a  circuit from~\cite{RahimiLPHD} (cf. Fig.~3 there) for generating hypervectors of trigrams (Section~\ref{sect:ngrams}) that used Barrel shifters to minimize the switching activity during the permutation operations.   
Note that the same circuit could have been designed using the hardware primitives for binding and permutation operations as the building blocks, but such a design would come at the price of the reduced efficiency.
Another common bottleneck in the hardware implementations of machine learning applications of HDC/VSA is the item memory (cf. Fig.~8 in~\cite{DattaProcessor2019}). 
The presence of this bottleneck caused researchers to consider ways of efficiently eliminating it.
A prominent way to do so is the rematerialization of the item memory using inexpensive recurrent methods as proposed in~\cite{SchmuckHardwareOptimizations2019, KleykoCA2020, AlwaysOn2021}. 
This idea of rematerialization created a room for trading off system's dynamic and leakage powers and was demonstrated to increase energy efficiency in scenarios involving, e.g., biosignal processing \cite{AlwaysOn2021,MenonWearableCA2021, MenonWearableCAEXT2022}.

In summary, we can argue that hardware implementations of HDC/VSA rely on architectural primitives corresponding to the basic elements and operations of HDC/VSA. 
However, in order to increase the efficiency, it is also common to design circuits implementing compositional computational primitives from Section~\ref{sect:primitives}.

\subsubsection{HDC/VSA combined with neural networks}
\label{sec:hardware:neural:nets}

The aforementioned works have demonstrated the benefits of HDC/VSA on relatively small-scaled classification tasks. In order to approach more complex tasks, a common strategy is to combine some of the basic HDC/VSA primitives (discussed in Section~\ref{sect:primitives}) with neural networks. 
For instance, representations from pretrained neural networks have been used with the HDC/VSA primitives to compactly represent a set of key-value pairs to generate image descriptors for visual place recognition~\cite{NeubertAggregation2021,NeubertPlaceRecognition2021}. 
One step further, the deep neural networks were trained from the scratch to be able to directly generate desired hypervectors that were further bound, or superposed by HDC/VSA operations to represent the concepts of interest~\cite{KarunaratneHDAugmented2020,FSCIL2022,herscheNVSA2022}. 
They achieved the state-of-the-art accuracy compared to the stand-alone deep learning solutions in various tasks involving images, including few-shot learning~\cite{KarunaratneHDAugmented2020},  continual learning~\cite{FSCIL2022}, and visual abstract reasoning~\cite{herscheNVSA2022}. The hardware implementation of such \emph{hybrid} architectures may vary. For instance, the associative memory for few-shot learning was implemented on the phase-change memory devices to execute searches in constant time, while the neural network was implemented externally~\cite{KarunaratneHDAugmented2020}. Alternatively, the whole architecture for the visual abstract reasoning was executed on CPUs, whereby leveraging HDC/VSA leads to two orders of magnitude faster execution than the functionally-equivalent symbolic logical reasoning~\cite{herscheNVSA2022}.

\section{Discussion}
\label{sect:related}

HDC/VSA has been criticized for lacking a structured methodology for designing systems as well as for missing well-defined design patterns ~\cite{Neubert2019}. 
Here (Section~\ref{sect:primitives}), we compiled existing computational primitives with HDC/VSA that paint a different picture. 
There is an HDC/VSA methodology addressing a wide range of applications, but it is scattered throughout the literature. 
In addition to compiling existing work, we laid out principles of design for building distributed representations of data structures such as sets, sequences, trees, key-value pairs, and more. 
This demonstrates a rich algorithmic and representation level approach which one can use as an abstraction for the next generation of computing devices.

Our compilation of varied HDC/VSA primitives also suggests that, contrary to some earlier assessments (see, e.g.,~\cite{VanderVeldeBlackboard2006} and the commentary in~\cite{GaylerJackendoff2006}), the repertoire potential of HDC/VSA applications is extremely wide, ranging from low-level sensory processing to high-level reasoning.  
While we provided an extensive introduction to HDC/VSA as well a comprehensive collection of computational primitives and existing connections to computing hardware, there was no goal to provide the complete state-of-the-art of the area such as, e.g., a review of all existing HDC/VSA models. 
We, however, hope that this article will motivate readers to explore the current state of the area that is covered in details in a two-part survey that covers  both fundamentals~\cite{KleykoSurveyVSA2021Part1} as well as applications~\cite{KleykoSurveyVSA2021Part2}.  
We think that the strength of HDC/VSA comes for the applications where there is a need for a computing framework  constructing transparent compositional distributed representations that will allow interfacing unconventional parallel computing hardware. 
It is not obvious how to achieve it with, e.g., modern neural networks, though it should be admitted that there is increasing empirical evidence demonstrating that certain problems benefit from hybrid approaches combining elements of HDC/VSA and neural networks.

That being said, it is still important to admit the limitations and challenges of HDC/VSA and, therefore, before ending the article, we would like to focus on them (Section~\ref{sect:disc:limit}). 
We conclude by discussing the role of HDC/VSA as a framework for computing with emerging hardware (Section~\ref{sec:related:framework}).

\subsection{Limitations and open challenges}
\label{sect:disc:limit}

Here, we would like to emphasize some of the limitations of HDC/VSA that are directly related to the scope of this article: applications (Section~\ref{sect:limt:app}), dimensionality of hypervectors (Section~\ref{sect:limt:dim}), and flow control (Section~\ref{sect:limt:flow}). 
For a broader discussion of open challenges, we kindly refer the reader to the section ``Open issues'' in~\cite{KleykoSurveyVSA2021Part2}.

\subsubsection{Applications}
\label{sect:limt:app}
There are numerous attempts to use HDC/VSA in problems within various application domains (see~\cite{KleykoSurveyVSA2021Part2} for a detailed coverage). 
Some well-known examples of using HDC/VSA include word embedding~\cite{KanervaRI2000,JonesMeaning2007} (though largely overshadowed after~\cite{MikolovWord2vec2013, pennington2014glove}), analogical reasoning~\cite{PlateStructure1997, RachkovskijSimilarity2012}, cognitive architectures~\cite{BuildBrain, RachkovskijWorldModel2013} and modeling~\cite{BlouwConcepts2016,KellyTesseract2017} as well as solving classification tasks~\cite{HDGestureIEEE, HDClassGe}.
It must be admitted, however, that most of these use-cases were limited to small scope problems, therefore, there is still a need to demonstrate how HDC/VSA-based solutions scale-up to real-world computational problems and, what is also important, to identify niches where the advantages of HDC/VSA will be self-evident.  
We think that further research will eventually address this limitation as we see two recent developments in this direction.
First, there is a continuing trend to extend HDC/VSA to novel domains -- promising recent examples include applications in communications~\cite{KimHDM2018} and in distributed systems~\cite{SimpkinHDWorkflow2019}. 
Second, there is an increasing number of studies (see Section~\ref{sec:hardware:neural:nets} and, e.g.,~\cite{KarunaratneHDAugmented2020,KleykoMANN2022, CheungSuperposition2019, herscheNVSA2022, NeubertAggregation2021,GanesanLearning2021}) that combine together neural networks and HDC/VSA primitives. 
This seems to be a promising way to scale-up HDC/VSA-based solutions to real-world problems, in the short-term.

\subsubsection{HDC/VSA dimensionality and working memory}
\label{sect:limt:dim}

The key feature of data representation in HDC/VSA is that data structures are represented by fixed sized hypervectors, independent of the size of a data structure. This is in contrast to the localist representations of data structures, that grow linearly or even quadratically with the number of elements. 
On the one hand, it is a great advantage as data structures of arbitrary size and shape can be manipulated in parallel with the elementary set of HDC/VSA operations. 
At the same time, as we have seen in Section~\ref{sect:primitives}, dimensionality of hypervectors might easily become a limitation since for a given dimensionality, the information content of representation, i.e., the HDC/VSA capacity, limits the size of data structures that can be represented reliably~\cite{Frady17,ThomasHDFoundations2020}.

Conceptually, one should think of the memory in hypervectors as the working memory or working registers, holding the data relevant during an ongoing computation. 
In contrast, the role of a long-term memory for a HDC/VSA-based system can be fulfilled by, e.g., a large capacity associative content-addressable memory that might store hypervectors of data structures~\cite{RachkovskijWorldModel2013, EmruliInteroperability2015}. 
Currently, this idea is being investigated by the community~\cite{Steinberg2022}.

The limitation of working memory in HDC/VSA has interesting parallels to the limitation of human working memory. 
For data structures of limited size, there are guarantees for exact reconstruction~\cite{ThomasHDFoundations2020}. However, transcending the theoretical bound for exact reconstruction, the data representation becomes lossy, with error rates also being theoretically predictable~\cite{Frady17}.
HDC/VSA representations of data structures in the lossy regime have been shown to reproduce some properties of human working memory. For example, the recall of sequence in an HDC/VSA, as described in Section~\ref{sect:seq}, can reproduce the performance of humans remembering sequences ~\cite{ChooSerialOrderRecall2010, KellyDeclarativeMemory2020}.
Further, the modeling of memorizing sequences with HDC/VSA was linked to the neuroscience literature in~\cite{CalmusBindingNeurobiologically2019}.  
It is not immediately clear how this capturing of the limitations of human memory might be beneficial in engineering applications. 
The way biological working memory coarsens its content and gradually degrades might be an important feature of cognition whose benefits are not yet fully appreciated.
However, for applications that require guarantees for exact reconstruction, the dimensionality of hypervectors needs to be specified at the design stage that makes it a limitation for the situations where data structures to be represented can be of highly varying size.

\subsubsection{Flow control}
\label{sect:limt:flow}

HDC/VSA implementations of algorithms generally rely on existing non-HDC/VSA mechanisms for flow control.
This is reasonable in systems where the aim is to use HDC/VSA to implement conventional computing approaches. 
In this case, it can be seen more from the point of extending conventional computing with HDC/VSA.
However, if we were modeling biological systems we should not be using non-HDC/VSA conventional computing flow control. 
Moreover, from the efficiency point of view, when using emerging hardware it might not be desirable to have a conventional processing unit for flow control.  
For these reasons, it is important to develop methodologies for flow control that would use native HDC/VSA primitives. 
In our opinion, it is possible. However, to date the efforts in this direction are quite limited. 
There was an attempt in~\cite{StewartSymbolicReasoning2010} to define a model of a biological system with HDC/VSA-based control. 
Two other related efforts are~\cite{UCBHD_FSA} that presented a proposal for a stack machine and~\cite{Laiho2015} proposing a processor with instructions specified in the form of hypervectors.

\subsection{HDC/VSA as a framework for computing with emerging hardware}
\label{sec:related:framework}

HDC/VSA was originally proposed in cognitive neuroscience as models for symbolic reasoning with distributed representations. More recently, it has been shown that HDC/VSA can formulate sub-symbolic computations, for example, in machine learning tasks. 

Here we proposed that HDC/VSA provides a computing framework within the algorithmic level of Marr's framework~\cite{MarrVision1982} for linking abstract computation and emerging hardware levels. 
The algorithmic formalism of HDC/VSA (with few exceptions) is the same for all of its variants.
Thus, HDC/VSA enables a model-independent formulation of computational primitives. 
At the same time, HDC/VSA also provides a seamless interface between algorithms and hardware. 
In Section~\ref{sect:frameworks}, we illustrated how different HDC/VSA models can connect to specific types of emerging hardware. 
Moreover, in Section~\ref{sect:comp:super} we demonstrated how HDC/VSA can be used for computing in superposition.
This feature extends HDC/VSA beyond the conventional computing architectures, and we foresee that together with algorithms that leverage computing in superposition, such as resonator networks~\cite{ResPart1, ResPart2} (Section~\ref{sec:resonator:network}), it will pave the way to efficient solutions of non-trivial combinatorial search problems (see examples in~\cite{FradyDisentangling2018,RennerUnderstanding2022}).

Another interesting aspect of computing with hypervectors is that it occupies a realm between
digital and analog computing. After each computation step in a digital computer, all vector components are pulled to one of the possible digital states (bits). 
This individual discretization of each component avoids error accumulation.  
Conversely, an analog computer is supposed to implement an analog dynamical system to predict its future states. 
Any deviation between the dynamical system to be analyzed and its computer implementation (e.g., noise) leads to uncontrollable error accumulation in analog computers. 
HDC/VSA operations leverage analog operations on vectors without discretization. 
However, discretization takes place on the entire vector level, when a resultant hypervector is matched with the entries in the item memory. 
Thus, HDC/VSA can leverage (potentially very) noisy dynamics in the high-dimensional state space of emerging hardware, while still protecting against error accumulation.

Despite all the above promising aspects, the practicability of the HDC/VSA computing framework for emerging computing hardware is yet to be thoroughly quantified.
An important future direction is to develop a systematic methodology to quantitatively measure and compare side-by-side the efficiency of different computing frameworks on a concrete hardware. In this article, we concentrated on the question how HDC/VSA enables   
the construction of varied algorithmic primitives and, therefore, could be a possible candidate framework in such a comparison.

\subsubsection*{Alternative frameworks}
\begin{table}[tb]
\renewcommand{\arraystretch}{1.0}
\caption{A qualitative assessment of HDC/VSA capabilities contrasted to conventional computing and neural networks.}
\label{fig:position}
    \begin{center}
    \begin{tabular}{c c c c} 
     &  \makecell{\textbf{Conventional} \\ \textbf{computing/AI}}  & \makecell{\textbf{Neural}\\ \textbf{networks}} & \textbf{HDC/VSA}  \\ \hline 
     \makecell{Distributed representation} & {\color{red} \xmark}  & {\color{cadmiumgreen} \cmark} & {\color{cadmiumgreen} \cmark} \\  \hline 
     \makecell{Learning from data } & {\color{red} \xmark}  & {\color{cadmiumgreen} \cmark} & {\color{cadmiumgreen} \cmark} \\  \hline 
     \makecell{Symbolic computing with \\ variables and binding} & {\color{cadmiumgreen} \cmark} & {\color{red} \xmark} & {\color{cadmiumgreen} \cmark} \\  \hline 
     \makecell{Tolerance to  \\ device imperfections} & {\color{red} \xmark}  & ? & {\color{cadmiumgreen} \cmark} \\  \hline 
     \makecell{Transparency} & {\color{cadmiumgreen} \cmark}  & {\color{red} \xmark} & {\color{cadmiumgreen} \cmark} \\  \hline 
     \end{tabular}
    \end{center}
\end{table}

HDC/VSA constitutes a computing framework that provides an algebraic language for formulating algorithms and, at the same time, links the computation to distributed states on hardware. 
Table~\ref{fig:position} compares the qualitative properties of HDC/VSA as a computing framework to conventional computing and neural networks. 

There is a tradeoff between how general a framework is in terms of computation and how closely it is linked to implementation. A general purpose framework typically requires a full sealing between implementation and computation, like, for example, the conventional computing architecture. Conversely, a framework that is well matched to an implementation, and, therefore, can efficiently leverage the hardware, is typically quite special purpose. We argue that the tradeoff HDC/VSA provides between generality and linking to implementation is ideal for emerging hardware. 
In particular, it seamlessly provides implementations of algorithms that leverage distributed representations, parallel operations, and can tolerate noise and imprecision~\cite{HDNP17}. 
Of course, HDC/VSA is not the only candidate of a framework for emerging hardware, alternative approaches include probabilistic computing~\cite{mansinghka2009natively},
sampling-based computing~\cite{orban2016neural}, computing by assemblies of neurons~\cite{PapadimitriouBrain2020}, and dynamic neural fields~\cite{schoner2016dynamic}.
For example, in neuromorphic computing, dynamic neural fields is an alternative computing framework that could support fully symbolic operations. In fact, dynamic neural fields and HDC/VSA might complement each other by combining real-time dynamics of dynamic neural fields with the computational power and scalability of HDC/VSA.  
The detailed comparison between these approaches and HDC/VSA is, however, outside the scope of this article. 
Nevertheless, in our opinion HDC/VSA is the most transparent approach in structuring computation, and the most general with regard to different types of hardware. In terms of formulating algorithms and computational primitives, HDC/VSA offers a common language, independent of a particular HDC/VSA model. 
For a desired computation on a given hardware, one of the many existing HDC/VSA models can provide the most advantageous implementation in terms of energy and time efficiency.

There is currently a plethora of collective-state computing approaches emerging, such as compressed sensing, Bloom filters, reservoir computing, etc.,  relying on distributed representations~\cite{CsabaCoupled2020}. These approaches are rather disjoint, and typically focus on special purpose computing applications.
HDC/VSA has been shown to be able to formalize different types of collective-state computing including reservoir computing~\cite{Frady17, intESN}, Bloom filters~\cite{ABF}, compressed sensing~\cite{FradySDR2020}, randomized kernels~\cite{FradyFunctions2021, FradyFunctionsNICE2022}, and extreme learning machines/random vector functional link networks~\cite{intRVFL}.
Thus, we see HDC/VSA as a promising candidate framework for providing a ``lingua franca'' of collective-state computing.


\appendices


\begin{table}[t]
    \centering
    \caption{Table of behaviour of (2,4) Turing machine.}
    \label{tab:24:turing}
    \begin{tabular}{| c| c | c |}
        \hline
         &  A & B  \\  \hline
         0 &  2 L A & 3 R A  \\  \hline
         1 &  3 L B & 2 L B  \\  \hline
         2 &  3 L A & 0 R B  \\  \hline
         3 &  3 L A & 1 R B  \\  \hline
    \end{tabular}
\end{table}	

\section{On Turing completeness of HDC/VSA}
\label{sect:turing}

\begin{table}[b]
    \centering
    \caption{The heteroassociative item memory implementing (2,4) Turing machine.}
    \label{tab:TM24:HA}
    \begin{tabular}{| c| c |c |c |}
        \hline
        Address (input) &  Content (output) & &  \\     
        \hline
         &  Tape content & Next State & Head's move  \\     
        \hline        
	    $\textbf{a} \odot \textbf{0}$ & $\textbf{2}$ & \textbf{a} & L  \\
        \hline   
        $\textbf{a} \odot \textbf{1}$ & $\textbf{3}$ & \textbf{b} & L  \\
        \hline
        $\textbf{a} \odot \textbf{2}$ & $\textbf{3}$ & \textbf{a} & L  \\
        \hline
        $\textbf{a} \odot \textbf{3}$ & $\textbf{3}$ & \textbf{a} & L  \\
        \hline
        $\textbf{b} \odot \textbf{0}$ & $\textbf{3}$ & \textbf{a} & R  \\
        \hline
        $\textbf{b} \odot \textbf{1}$ & $\textbf{2}$ & \textbf{b} & L  \\
        \hline
        $\textbf{b} \odot \textbf{2}$ & $\textbf{0}$ & \textbf{b} & R  \\
        \hline
        $\textbf{b} \odot \textbf{3}$ & $\textbf{1}$ & \textbf{b} & R  \\
        \hline
    \end{tabular}
\end{table}	

It is practical to have a collection of primitives for common data structures. However, these primitives alone do not provide us with a quantification of the theoretical capabilities of using HDC/VSA as a computing framework.  
Of course, it is desirable that a computing framework for emerging hardware be able to (in theory, at least) execute any algorithm.
For example, in~\cite{zhang2020system} that proposed a system hierarchy for neuromorphic computing, it has been emphasized that Turing completeness is an essential property for an abstraction model that is used at the algorithmic level.  
Therefore, in this section, we sketch ways of demonstrating that  
HDC/VSA is computationally universal by  exemplifying  how they (with some assumptions) can be used to emulate systems that have already been proven to be Turing complete.
While computing in superposition is likely to be the most interesting feature of operating with HDC/VSA, computational universality is still a critical property to study as it characterizes the general computational power of a system. 
It is worth noting that among HDC/VSA researchers there is a general agreement that HDC/VSA are computationally universal but, to the best of our knowledge, this has not been shown yet. 
Therefore, here we make two proposals towards demonstrating their universality: by implementing a Turing machine and by emulating  an elementary cellular automaton, which is also known to be Turing complete~\cite{Cook2004}.
Note that while these proposals might not be tight enough to be qualified as a formal proof, we believe that the directions below are the most promising ways to make such a proof.

\subsection{Implementation of Turing machines with HDC/VSA}
\label{sect:turing:TM}

Since there are a number of small Turing machines known to be universal~\cite{NearyUniversal2009}, we first focus on demonstrating how HDC/VSA can be used as a part of an implementation of such a machine. 
In order to do so, we present how HDC/VSA representations are used to map a table of behaviour~\cite{NearyUniversal2009} and execute the machine.

The presented implementation could be used to realize any Turing machine, but for the sake of compactness we exemplify the implementation with a (2,4) Turing machine, which has 2 states (A and B) and 4 symbols (0, 1, 2, 3).
The table of behaviour of a (2,4) Turing machine is presented in Table~\ref{tab:24:turing}. 
For a given combination of the current state and tape's content, it provides which symbol should be written to the current cell, the next state of the machine, and the direction for the head's move.

\subsubsection{HDC/VSA implementation of the table of behaviour}

\begin{figure}[tb]
\centering
\includegraphics[width=1.0\columnwidth]{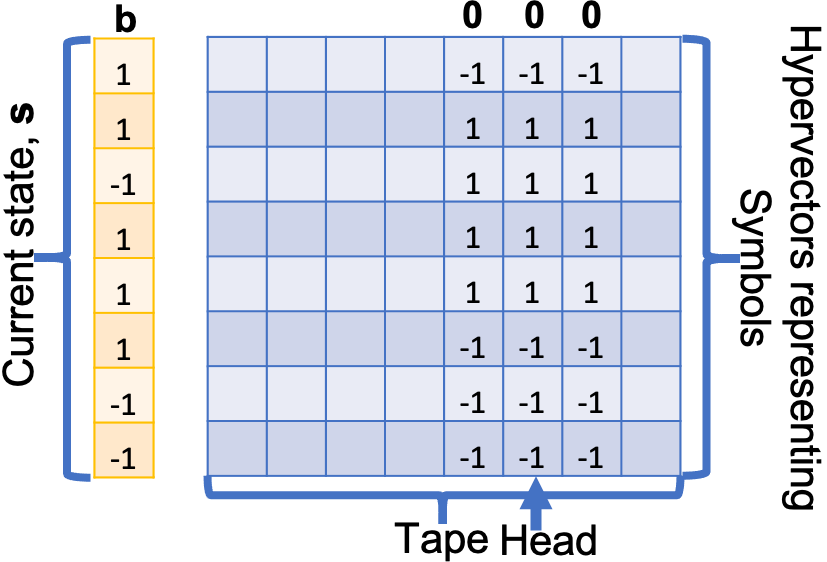}
\caption{
An illustration of the current state of the machine and its tape. 
}
\label{fig:tape:1}
\end{figure}

We use the Multiply-Add-Permute model described above. 
In order to represent the table of behaviour of a Turing machine, we first create two item memories populated with random hypervectors.
One item memory stores the states, e.g., in the case of a (2,4) Turing machine it includes only two hypervectors for states A and B (denoted as $\textbf{a}$ and $\textbf{b}$), respectively. 
Another item memory stores hypervectors for symbols.
Since the considered machine uses only four symbols, four hypervectors $\textbf{0}$, $\textbf{1}$, $\textbf{2}$, and $\textbf{3}$ are sufficient. 
These item memories are used to construct a hypervector for each combination of states and symbols.  
The hypervector is constructed by applying the binding operation on the hypervectors for a state and a symbol.

Eight hypervectors corresponding to all possible combinations  form a basis for constructing a third, heteroassociative, item memory, i.e., the memory where the address and content parts store different hypervectors. 
The heteroassociative item memory can implement any table of behaviour by using the bound pair of state and symbol as input to the memory and issuing hypervectors, which should be used as the tape content, head's move, and next state as an output.      
Table~\ref{tab:TM24:HA} presents the heteroassociative item memory for the table of behaviour of (2,4) Turing machine.
Thus, three item memories constitute the static part of the system, which is generated only once at the initialization. 
At this point, it is worth making a note that in addition to the standard assumptions about unlimited time and memory resources, there is an extra assumption about the heteroassociative item memory.
In particular, it should be guaranteed to behave correctly in the absence of external errors. 
Practically, it means that the address part of the heteroassociative item memory should not have repeated entries. 
Even for moderate dimensionality of hypervectors the chance of such an event is low, but if this happens the issue is solved by the regeneration of the item memories.

\begin{figure}[tb]
\centering
\includegraphics[width=1.0\columnwidth]{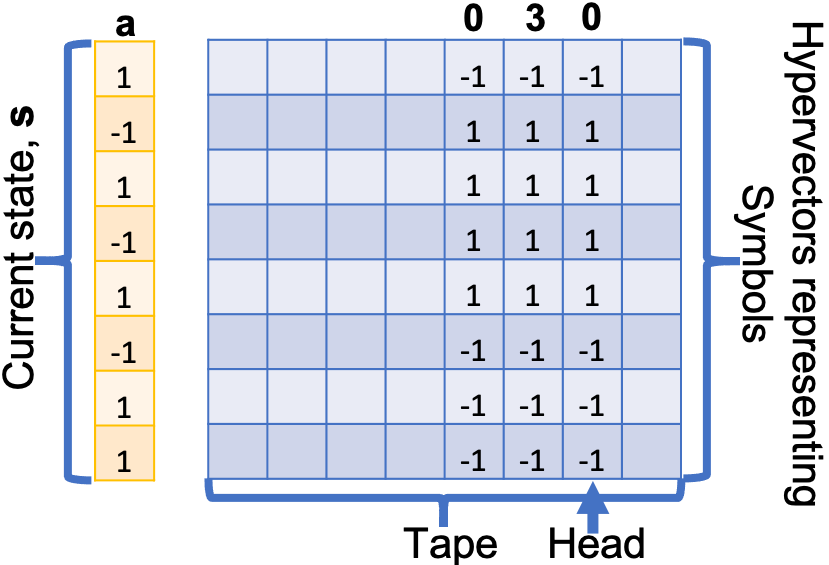}
\caption{
The updated state and tape of the machine after the previous state as in Fig.~\ref{fig:tape:1}.
}
\label{fig:tape:2}
\end{figure}

\subsubsection{HDC/VSA-based tape}

The other part of the system is dynamic and includes the location for storing a hypervector for the current state, the tape, and current position of the head. 
Fig.~\ref{fig:tape:1} presents an example of the dynamic part of the system. 
In the case of using HDC/VSA, the tape can be seen as a matrix where each column corresponds to the hypervector of a symbol.
In order to make the next step, the machine has to read the hypervector of the current state ($\textbf{b}$ in Fig.~\ref{fig:tape:1}) and the hypervector of the symbol at the current location of the head ($\textbf{0}$ in Fig.~\ref{fig:tape:1}). 
The result of binding of these hypervectors $\textbf{b} \odot \textbf{0}$ is used as an input to the heteroassociative memory.
The output of the memory indicates that hypervector $\textbf{a}$ should be written to the current state; the tape's content is changed to $\textbf{3}$, and the head should be moved to the right of the current location. 
The updated state is shown in Fig.~\ref{fig:tape:2}.
In such a manner, the system could operate on the tape for the required number of computational steps. 
Summarizing, the proposed implementation of a Turing machine uses basic elements of HDC/VSA such as hypervectors, item memories, and the binding operation; however, it also includes few parts that go beyond HDC/VSA -- namely, control of head movements and unlimited memory tape.

\subsubsection{Scaling HDC/VSA implementation}

\begin{figure}[tb]
\centering
\includegraphics[width=1.0\columnwidth]{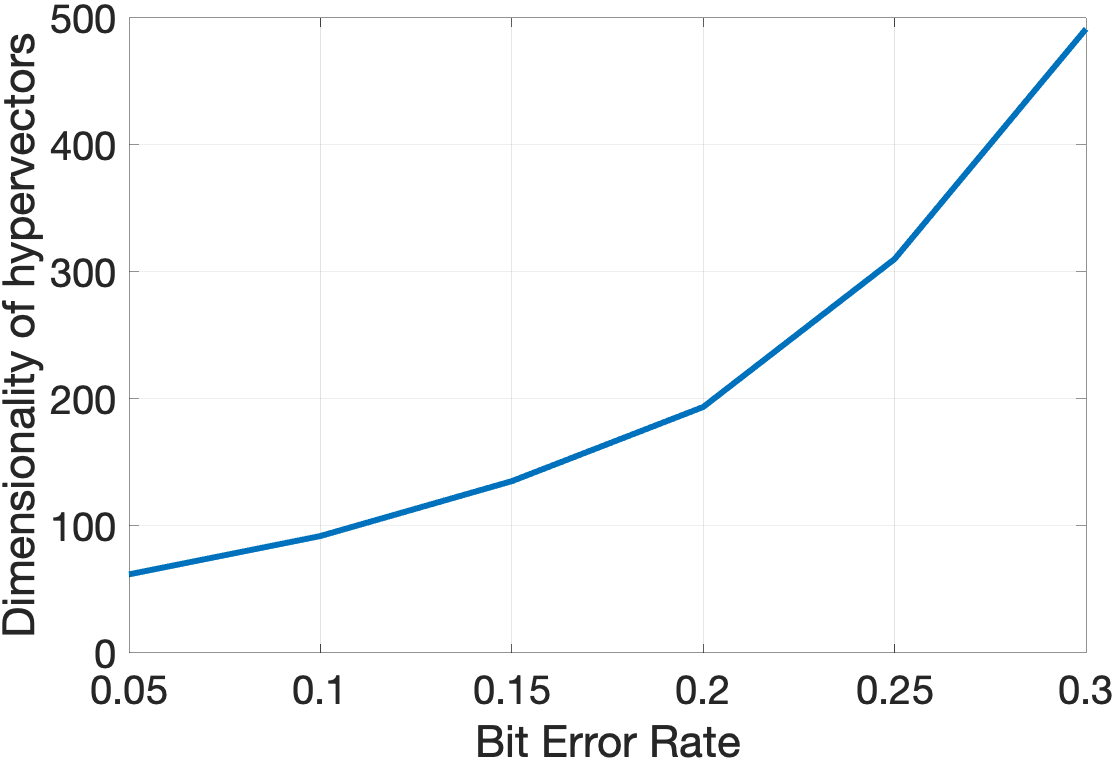}
\caption{
The average dimensionality of hypervectors required to make at least $10^9$error-free updates of the emulated (2,4) Turing machine when the hypervectors representing symbols on the tape were subject to external bit flips. 
The Bit Error Rate was in the range $[0.05, 0.30]$ with step $0.05$.
The results were computed from $10$ simulation runs with random initializations of hypervectors in the item memories and random bit flips added at every update of the machine.
}
\label{fig:TM:vsas}
\end{figure}

Since the proposed implementation of a Turing machine does not make use of the superposition operation, there is no crosstalk noise being introduced to the computations, which in turn means that in the absence of external noise the emulation behaves in a deterministic way.
Thus, even tiny three-dimensional vectors can be used to construct the heteroassociative item memory with unique entries. 
Nevertheless, since one of the arguments in favour of HDC/VSA is their built-in tolerance to errors, it is interesting to observe the behaviour of the emulation in the presence of external noise. 
We performed simulations where the external noise was added to the tape by randomly flipping signs of a fraction of hypervector components.  
Fig.~\ref{fig:TM:vsas} presents the average dimensionality of hypervectors required to make at least $10^9$ error-free updates of the emulated Turing machine when the hypervectors representing symbols on the tape were subject to external bit flips. 
The Bit Error Rate varied in the range $[0.05, 0.30]$ with step $0.05$.
The starting dimensionality of hypervectors was $2^4$.
If the error in emulation was happening in less than $10^9$ 
steps, then the dimensionality was increased by $10$\%.
The results demonstrate that the proposed implementation can reliably emulate the Turing machine given adequate resources (i.e., dimensionality of hypervectors).
Naturally, in the presence of external noise, more resources are needed to obtain the error-free execution of the machine.
Nevertheless, an important observation is that the implementation works with imprecise noisy representations.
Moreover, the robustness of the implementation comes at no cost in terms of design, as the same algorithm is being used for any amount of noise and the only cost to be paid is the increased size of the system.

\begin{figure}[tb]
\centering
\includegraphics[width=1.0\columnwidth]{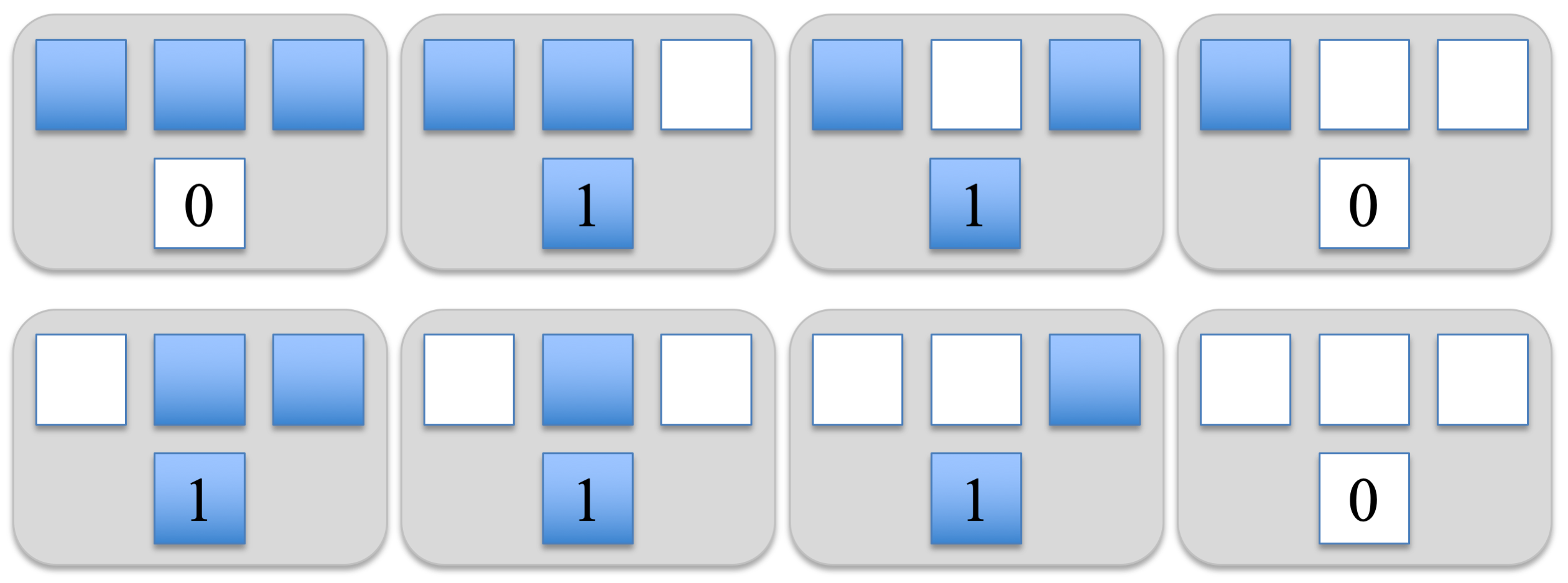}
\caption{The assignment of new states for a center cell when the cellular automaton uses rule 110. A hollow cell corresponds to zero state while a shaded cell marks one state.
}
\label{fig:ca110}
\end{figure}

\subsection{Emulation of cellular automaton with HDC/VSA}
\label{sect:turing:CA}

Since HDC/VSA are designed to create vector representations of symbolic structures, when identifying a Turing complete system suitable for emulation with HDC/VSA, it is also natural to choose a highly structured system which uses a small finite alphabet of symbols.   
We think that an elementary cellular automaton is one example of such a system. 
Since the elementary cellular automaton with the rule $110$ is known to be Turing complete~\cite{Cook2004}, we would like to demonstrate how HDC/VSA can be used in emulating this rule. 
In order to do so, we first revisit the elementary cellular automaton concept.  
Next, we present a HDC/VSA algorithm for mapping and executing an elementary cellular automaton.
Thus, we literally follow the roadmap from~\cite{Cook2004}: ``The automaton itself is so simple that its universality gives us a new tool for proving that other systems are universal''. 
Finally, we explore how the proposed implementation is scaling with respect to the size of the initial grid state of an elementary cellular automaton, the dimensionality of hypervectors, and the amount of noise present during the computations. 
The major point of the latter is that even for large amount of noise the implementation can perfectly emulate the elementary cellular automaton given sufficiently large dimensionality of hypervectors, which is a nice property as the robustness is achieved without modifying the design.

\subsubsection{Elementary cellular automata}

An elementary cellular automaton is a discrete  computational model consisting of a one-dimensional grid of cells~\cite{Wolfram}. 
Each cell can be in one of a finite number of states (two -- for the elementary automaton). 
States of cells evolve in discrete time steps according to a fixed rule. 
The state of a cell at the next  computational step depends on its current state and states of its neighbors.  
The computations performed by an elementary cellular automaton are local.  
The new state of a cell is determined by previous states of the cell itself  and its two
neighboring cells (left and right). 
Thus, only three cells are involved in a computation step, i.e., for binary states, there are in total $2^3=8$ combinations.  
A rule assigns states for each of eight combinations.  
Fig.~\ref{fig:ca110} presents all combinations and the corresponding states for the rule $110$.

\subsubsection{HDC/VSA algorithm for emulating an elementary cellular automaton with the rule 110}

\begin{table}[t!]
    \centering
    \caption{The heteroassociative item memory implementing rule $110$.}
    \label{tab:ca:rule110}
    \begin{tabular}{| c| c |}
        \hline
        Address (input) &  Content (output)  \\     
        \hline
	$\textbf{h}_{111}=  [  \textbf{l} \odot \textbf{1} +  \textbf{c} \odot \textbf{1}  + \textbf{r}  \odot \textbf{1}  ]$ & $\textbf{0}$  \\
        \hline      
	$\textbf{h}_{110}=  [  \textbf{l} \odot \textbf{1} +  \textbf{c} \odot \textbf{1}  + \textbf{r}  \odot \textbf{0}  ]$ & $\textbf{1}$  \\
        \hline  
	$\textbf{h}_{101}=  [  \textbf{l} \odot \textbf{1} +  \textbf{c} \odot \textbf{0}  + \textbf{r}  \odot \textbf{1}  ]$ & $\textbf{1}$  \\
        \hline  
	$\textbf{h}_{100}=  [  \textbf{l} \odot \textbf{1} +  \textbf{c} \odot \textbf{0}  + \textbf{r}  \odot \textbf{0}  ]$ & $\textbf{0}$  \\
        \hline  
	$\textbf{h}_{011}=  [  \textbf{l} \odot \textbf{0} +  \textbf{c} \odot \textbf{1}  + \textbf{r}  \odot \textbf{1}  ]$ & $\textbf{1}$  \\
        \hline  
	$\textbf{h}_{010}=  [  \textbf{l} \odot \textbf{0} +  \textbf{c} \odot \textbf{1}  + \textbf{r}  \odot \textbf{0}  ]$ & $\textbf{1}$  \\
        \hline  
	$\textbf{h}_{001}=  [  \textbf{l} \odot \textbf{0} +  \textbf{c} \odot \textbf{0}  + \textbf{r}  \odot \textbf{1}  ]$ & $\textbf{1}$  \\
        \hline  
	$\textbf{h}_{000}=  [  \textbf{l} \odot \textbf{0} +  \textbf{c} \odot \textbf{0}  + \textbf{r}  \odot \textbf{0}  ]$ & $\textbf{0}$  \\
        \hline  
    \end{tabular}
\end{table}	

We use the Multiply-Add-Permute model described above. 
In order to represent an elementary cellular automaton with the rule $110$, we first create two item memories populated with random hypervectors.
One item memory stores the finite alphabet, i.e., it includes only two hypervectors, for one and for zero (denoted as $\textbf{1}$ and $\textbf{0}$, respectively). 
Another item memory stores hypervectors for positions.
Since an elementary cellular automaton relies only on a cell in focus and its immediate neighbors, then three hypervectors: $\textbf{l}$ (left), $\textbf{c}$ (center), and $\textbf{r}$ (right) are sufficient. 
These item memories are used to construct a hypervector for each combination of states in three consecutive cells.  
The hypervector is constructed by applying the superposition operation on the bound pairs of a positional hypervector and an alphabet hypervector. 
In other words, the current states in three consecutive cells are represented as a set of unordered pairs. 
For example, for $010$, the corresponding compound hypervector is constructed as:
\noindent
\begin{equation*}
\textbf{h}_{010}=  [  \textbf{l} \odot \textbf{0} +  \textbf{c} \odot \textbf{1}  + \textbf{r}  \odot \textbf{0}  ].
\label{eq:ca110}
\end{equation*}
\noindent
All eight compound hypervectors form a basis for constructing a heteroassociative item memory which can implement any elementary rule by using the compound hypervectors as input to the memory, and issuing either $\textbf{1}$ or $\textbf{0}$ (determined by the rule) as an output.      
Table~\ref{tab:ca:rule110} presents the heteroassociative item memory for the rule $110$.
Thus, three item memories constitute the static part of the system, which is generated only once at the initialization.

The other part of the system performs computations for a given initial grid state of length $l$ at time $t=0$.
The initial grid state is mapped to a compound hypervector (denoted as $\textbf{a}_0$).
The mapping is done by applying the superposition operation on all hypervectors representing the states of cells at all positions.
Position $j$ in the grid is represented by applying the permutation operation $j$ times to the hypervector corresponding to a state at position $j$.
Thus, this representation corresponds to the mapping of a sequence with the superposition operation.
For example, if the initial grid state is $10101$, then  the representation of the state at the fifth position is $\rho^5 \textbf{1}$ while the compound hypervector for the initial grid state is:
\noindent
\begin{equation*}
\textbf{a}_{0}=  [ \rho^1 \textbf{1} +  \rho^2 \textbf{0} + \rho^3 \textbf{1} + \rho^4 \textbf{0} + \rho^5 \textbf{1} ].
\label{eq:ca110:grid}
\end{equation*}
\noindent

Given $\textbf{a}_{0}$, the next step is to compute $\textbf{a}_{1}$ or in general compute $\textbf{a}_{t+1}$ given $\textbf{a}_{t}$.

\begin{figure}[tb]
\centering
\includegraphics[width=1.0\columnwidth]{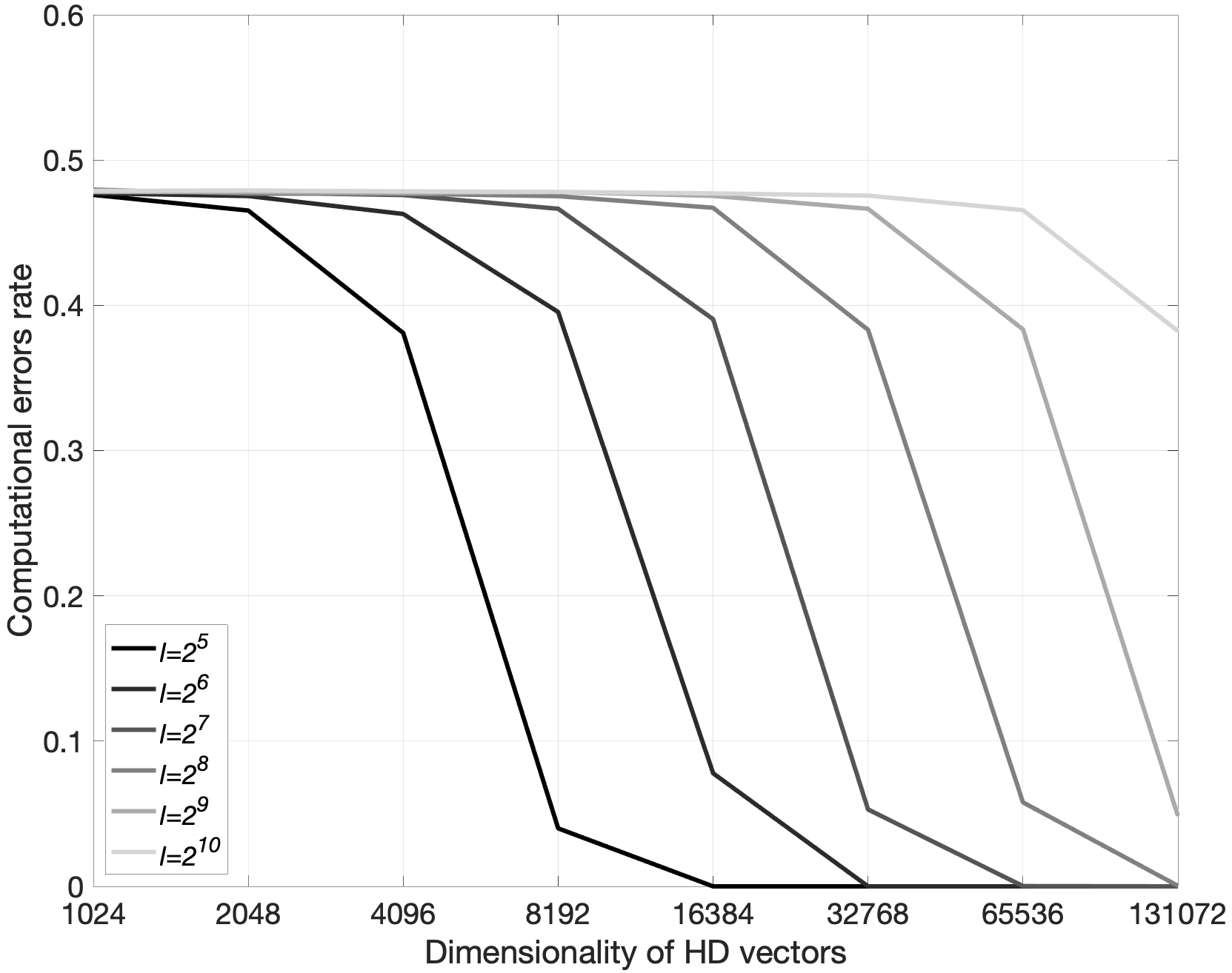}
\caption{
The average error rate after $100$ computational steps of the elementary cellular automaton against the dimensionality of hypervectors ($N=2^i$, $i \in [10, 17]$) for several different lengths of the grid ($l=2^i$, $i \in [5, 10]$).
The results were computed from $100$ simulation runs with random initializations of hypervectors in the item memories.
The initial grid states were also randomized.
}
\label{fig:ca110:vsas}
\end{figure}

First, $\textbf{a}_{t+1}$ is initialized to be an empty hypervector.
Next, for each position $j$ ranging from $1$ to $l$ we do the following (this step can be either serial or parallel):
\begin{itemize}

\item Approximately recover the state at $j$ and its neighbors as  $\hat{\textbf{h}} = [  \textbf{l} \odot \rho^{-(j-1)} \textbf{a}_{t} +  \textbf{c} \odot \rho^{-j} \textbf{a}_{t}  + \textbf{r}  \odot \rho^{-(j+1)} \textbf{a}_{t}  ]$.

\item Use $\hat{\textbf{h}}$ as the query to the heteroassociative item memory. The memory returns the content (i.e., $\textbf{0}$ or $\textbf{1}$) for the address closest to $\hat{\textbf{h}}$ in terms of dot product. The returned content is denoted as $\textbf{v}_j$.

\item Modify $\textbf{a}_{t+1}$ with $\textbf{v}_j$ as: $\textbf{a}_{t+1} += \rho^j \textbf{v}_j$.

\end{itemize}

Finally, apply the majority rule on $\textbf{a}_{t+1}$: $\textbf{a}_{t+1} = [\textbf{a}_{t+1}]$, so that it becomes bipolar.
In such a manner, the system could iterate through the grid for the required number of computational steps.

Last but not least, it is worth explicating that the proposed implementation assumes parts that go beyond HDC/VSA. First, the full computational system has its control architecture that is responsible for initializing the grid state as well as for running the for-loop, which can be seen as a recurrent connection, required for constructing $\textbf{a}_{t+1}$.
The second part that is assumed here to be the same as in the standard implementation of a cellular automaton, is the circuit determining when to stop the computation.
We have not focused on this circuit as our main goal here was to demonstrate how to evolve HDC/VSA representations to perform cellular automaton's computations.

\subsubsection{Scaling HDC/VSA emulation}

\begin{figure}[tb]
\centering
\includegraphics[width=1.0\columnwidth]{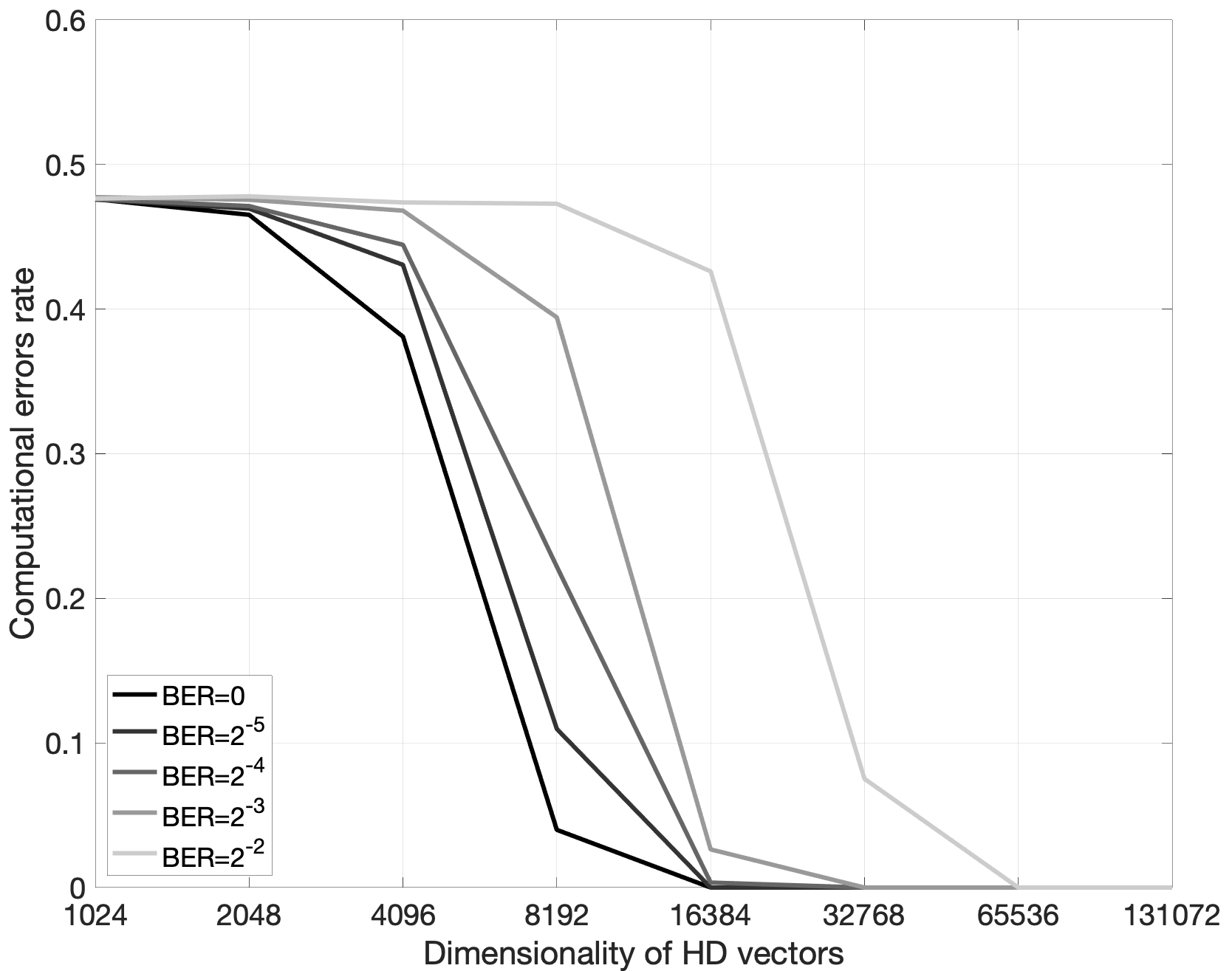}
\caption{
The average error rate after $100$ computational steps of the elementary cellular automaton against the dimensionality of hypervectors ($N=2^i$, $i \in [10, 17]$) for several different bit error rates, BER ($p=2^{-i}$, $i \in [2, 5]$) for the length of the grid $l=32$.
The results were computed from $100$ simulation runs with random initializations of hypervectors in the item memories.
The initial grid states were also randomized.
}
\label{fig:ca110:vsas:noise}
\end{figure}

It is known that compound hypervectors can be used to retrieve their components (Section~\ref{sect:probing}); however, there is a limit on the number of components which can be stored in a compound hypervector without losing the ability to recover the components~\cite{Frady17}. 
The rule of thumb is that for larger hypervector dimensionalities more components can be recovered from a compound hypervector.
For the task of emulating an elementary cellular automaton, it is important that $\hat{\textbf{h}}$ is similar enough to the correct state hypervector in the item memory.
Otherwise, we will introduce errors to the computations being emulated, which is highly undesirable. 
When constructing $\hat{\textbf{h}}$,  the main source of noise is the crosstalk noise from other cell states stored in $\textbf{a}_{t}$.
Therefore, in order to avoid errors in the computations, the dimensionality of hypervectors should depend on the length of the grid: the longer is the grid, the larger dimensionality is required for robustly querying the item memory.\footnote{In principle, it should be possible to analytically find the minimal dimensionality of hypervectors for robustly emulating the grid of the given length.}

Fig.~\ref{fig:ca110:vsas} presents the empirical results for a range of $l$ and $N$ values. 
The curves depict the average error rate after $100$ computational steps of the elementary cellular automaton.
Note that the errors occurring at the earlier computational steps will most likely propagate to the successive steps.
The length of the grid, $l$, varied as $2^i$, $i \in [5, 10]$, while the dimensionality of hypervectors, $N$, varied as $2^i$, $i \in [10, 17]$.
Thus, the results demonstrate that HDC/VSA can perfectly emulate the elementary cellular automaton with the grid of certain length, given adequate resources (i.e., dimensionality of hypervectors).

Note that Fig.~\ref{fig:ca110:vsas} presented the results for the case when hypervectors did not include any external noise. 
Since one of the arguments in favour of HDC/VSA is their built-in tolerance to errors, it is interesting to observe the behaviour of the emulation in the presence of external noise.
External noise was added by randomly flipping a fraction of components in $\textbf{a}_{t}$, but it was still assumed that the control architecture functions without errors. 
Fig.~\ref{fig:ca110:vsas:noise} presents the average error rate after $100$ computational steps of the elementary cellular automaton in the presence of external noise.
The bit error rate, $p$, varied as $2^{-i}$, $i \in [2, 5]$.
The length of the grid was fixed to $l=32$.

The results demonstrate that, naturally, in the presence of external noise, more resources are needed to obtain the error-free emulation.
Nevertheless, an important observation is that the HDC/VSA-based system works with imprecise noisy representations.
Moreover, the robustness of the system comes at no cost in terms of design, as the same algorithm is used in both cases and the only cost to be paid is in the increased size of the system.

\subsubsection{Studies related to computational universality of HDC/VSA}

Studying computational universality of a particular computing framework is important for understating ultimate theoretical limitations of computing hardware using this framework. 
For example, \cite{SIEGELMANN1991} has shown that recurrent neural networks are computationally universal,  \cite{PerezCompleteness2019} has shown universality of modern transformer and Neural GPU networks. 
Since HDC/VSA can express some recurrent neural networks~\cite{intESN}, studying their universality by leveraging on the existing results for neural networks is a possible direction of research. 
We, however, followed earlier approaches that showed that neural network-like systems can implement Turing machines~\cite{beimGraben2012}.  
In the sections above, we sketched how HDC/VSA can be used in implementations of a small Turing machine~\cite{NearyUniversal2009} and a universal elementary cellular automaton with the rule $110$~\cite{Cook2004}.

Recently, \cite{Kwisthout20} emphasized the need for a formal machine model for novel neuromorphic hardware in order to develop a computational complexity theory for neuromorphic computations.
This is an important direction of research for understanding the full potential of emerging hardware.
They argued, however, that in order to encompass the computational abilities of neuromorphic hardware, one will likely need to define an entirely new computing theory framework.
Their study has proposed to use spiking neural networks (shown to be Turing complete~\cite{Maass1996}) because, similar to HDC/VSA, they are suitable for co-located computation and memory, and massive parallelism -- which is not the case for the conventional computing architecture.

In addition to the demonstration of universality, an important practical question is how a complete computational architecture should look like. This is still an open question. 
A proposal has been sketched in~\cite{Laiho2015}, which featured a HDC/VSA-based processor where both data and instructions were represented as hypervectors.
There is another approach known as Tensor Product Variable Binding, which is closely related to HDC/VSA.
For example, Tensor Product Variable Binding can also be used to represent data structures in distributed representations~\cite{demidovskij2021encoding}.
The study~\cite{Smolensky1990} has demonstrated how to implement push, pop, and the Lisp primitives CAR and CDR with Tensor Product Variable Binding, while~\cite{Smolensky1989} has demonstrated how to implement a production system.
A HDC/VSA-based model, which was positioned as a general-purpose neural controller playing a role analogous to a production system, was proposed in~\cite{StewartSymbolicReasoning2010}.

Another relevant result is a demonstration of the feasibility of implementing Fluid Construction Grammars with HDC/VSA~\cite{KnightGrammar2015}.
Even though Fluid Construction Grammars have not been shown to be universal, it is a powerful and interesting approach for both cognitive and evolutionary linguistics.  \cite{KnightGrammar2015} proposed a vision similar to the one presented in Fig.~\ref{fig:stack}. They suggest  HDC/VSA can be seen as a ``virtual machine'' that can have different (independent) physical implementations, such as an indirect mapping to spiking neurons~\cite{TPAM} or direct mapping of operations with analog/digital implementations~\cite{MemristorHD19}.

\newcommand{\mb}{\mathbf}

\section{Summary of Vector-Symbolic Space and Operations}
\label{sect:VSA:summary}

\subsection{Key components}

This appendix presents excerpts from Section~\ref{sect:methods}
providing a summary of HDC/VSA.  The key components of all HDC/VSA are:
\noindent
\begin{itemize}
    \item High-dimensional space (e.g., bipolar);
    \item Orthogonality;
    \item Similarity measure (e.g., dot product $\langle \mb{a},\mb{b}
      \rangle$);
    \item Seed representations (e.g., random i.i.d. vectors);
    \item Operations on representations.
\end{itemize}

\noindent
There are three key operations in HDC/VSA:
\noindent
\begin{itemize}
    \item Binding (denoted as $\odot$, implemented as
      component-wise multiplication (Hadamard product) in the
      Multiply--Add--Permute model);
    \item Superposition (denoted as $+$, implemented as component-wise
      addition, enclosed in [\ldots] when thresholded);
    \item Permutation (denoted as $\rho$, e.g., rotation of
      coordinates).
\end{itemize}

\noindent
Below, we present the properties of the implementations of these
operations for the Multiply--Add--Permute HDC/VSA model~\cite{MAP}.  Here,
we enumerate the properties assuming that the seed hypervectors are
bipolar.

\subsection{Properties of the binding operation}

\begin{itemize}
    \item Binding is commutative: $\mb{a} \odot \mb{b} = \mb{b}
      \odot \mb{a}$;
    \item Binding distributes over superposition: $\mb{c} \odot
      (\mb{a} + \mb{b}) = \mb{c} \odot \mb{a} + \mb{c} \odot \mb{b}$;
    \item Binding is invertible: $(\mb{a} \odot \mb{b}) \odot
      \mb{b} = \mb{a}$ (bipolar $\mb{b}$ is self-inverse),  the inverse operation is
      called releasing or unbinding;
    \item Binding is associative: $(\mb{a} \odot \mb{b}) \odot
      \mb{c} = \mb{a} \odot (\mb{b} \odot \mb{c})$;
    \item The result of binding is dissimilar to each of its
      argument hypervectors: $\langle(\mb{a} \odot \mb{b}), \mb{a}
      \rangle \approx \langle (\mb{a} \odot \mb{b}), \mb{b} \rangle
      \approx 0$, hence binding is a ``randomizing'' operation;
    \item Binding preserves similarity: $\langle (\mb{c} \odot
      \mb{a}),(\mb{c} \odot \mb{b}) \rangle$ = $\langle \mb{a},\mb{b}
      \rangle$.
\end{itemize}

\subsection{Properties of the superposition operation}

\begin{itemize}
    \item Superposition is invertible: $(\mb{a} + \mb{b}) + (-\mb{b}) =
      \mb{a}$; for thresholded superposition: $\langle[[\mb{a} + \mb{b}] +
        (-\mb{b})],\mb{a} \rangle > 0$;
    \item In contrast to binding and permutation operations,
      the result of superposition $\mb{z}=\mb{a} + \mb{b}$ (often called the superposition
      hypervector) is similar to each of its argument hypervectors:
      i.e., the dot product between $\mb{z}$ and $\mb{a}$ or $\mb{b}$
      is considerably greater than $0$, $ \langle \mb{z}, \mb{a} \rangle \gg 0$ and
      $\langle \mb{z}, \mb{b} \rangle \gg 0$;
    \item Superposition is commutative: $\mb{a} + \mb{b} = \mb{b} +
      \mb{a}$;
    \item Thresholded superposition is approximately associative: $[[\mb{a}
        + \mb{b}] + \mb{c}] \approx [\mb{a} + [ \mb{b} + \mb{c}]] $.
\end{itemize}

\subsection{Properties of the permutation operation}

\begin{itemize}
    \item Permutation is invertible: $\rho^{-1}(\rho(\mb{a})) =
      \mb{a}$;
    \item Permutation distributes over both
      binding and superposition: $\rho (\mb{a} \odot \mb{b}) = \rho
      (\mb{a}) \odot \rho (\mb{b})$ and $\rho (\mb{a} + \mb{b}) = \rho
      ( \mb{a}) + \rho ( \mb{b})$;
    \item Similar to the binding operation, a random
      permutation $\rho$ results in a vector that is dissimilar to the
      argument hypervector: $\langle \rho(\mb{a}),\mb{a} \rangle
      \approx 0$, hence permutation is a ``randomizing'' operation;
      \item Permutation preserves similarity: $\langle \rho(\mb{a}),
   \rho(\mb{b}) \rangle = \langle \mb{a},\mb{b} \rangle$.
\end{itemize}

\section*{Acknowledgement}
The authors thank members of the Redwood Center for Theoretical Neuroscience and Berkeley Wireless Research Center for stimulating discussions. 
We would also like to thank Ross W. Gayler and Sohum Datta for their in depth comments on the early versions of this article. 
Finally, we would like to thank three anonymous reviewers and the editors for their insightful feedback as well as Linda Rudin for the careful proofreading that contributed to the final shape of the article.

\bibliographystyle{IEEEtran} 
\bibliography{references}

\end{document}